\documentclass[acmsmall,screen]{acmart}
\AtBeginDocument{%
  }

\setcopyright{acmlicensed}
\copyrightyear{2024}
\acmYear{2024}
\acmDOI{XXXXXXX.XXXXXXX}

\usepackage{graphicx} 
\usepackage{xspace}
\usepackage{pgfplots}
\usepackage{pifont}
\usepackage{multirow}
\usepackage{tcolorbox}
\pgfplotsset{compat=1.18}
\usepgfplotslibrary{groupplots}
\usepackage{soul}
\usepackage{hyperref}
\usepackage{multirow}
\usepackage{listings}
\usepackage{fancybox}
\usepackage{enumitem}
\usepackage{graphicx}
\usepackage{color}
\usepackage{array}
\usepackage{amsmath}
\usepackage{xspace}
\usepackage{url}
\usepackage{tikz}
\usepackage{caption}
\usepackage{pgfplots}
\usepackage{balance}
\usepackage{subcaption}
\pgfplotsset{compat=1.16}
\usepackage{pgf-pie}
\usetikzlibrary{pgfplots.statistics,calc}
\usepackage{algorithm}
\usepackage{algorithmic}
\usepackage{adjustbox}
\usepackage{makecell}

\usepackage{tcolorbox}
\makeatletter
\newcommand{\mybox}[1]{%
	\setbox0=\hbox{#1}%
	\setlength{\@tempdima}{\dimexpr\wd0+13pt}%
	\begin{tcolorbox}[boxrule=0.5pt, colback=white, arc=4pt,
		left=6pt,right=6pt,top=6pt,bottom=6pt,boxsep=0pt]
		#1
	\end{tcolorbox}
}

\definecolor{songcolor}{RGB}{191,191,191}

\newcommand{\tool}{PET-Select\xspace}
\newcommand*\circled[1]{\tikz[baseline=(char.base)]{
            \node[shape=circle,draw,inner sep=2pt] (char) {#1};}}

\title{Selection of Prompt Engineering Techniques for Code Generation through Predicting Code Complexity}
\author{Chung-Yu Wang}
\affiliation{
  \institution{York University}
  \city{Toronto}
  \country{Canada}
}
\email{cywang14@yorku.ca}

\author{Alireza DaghighFarsoodeh}
\affiliation{%
  \institution{York University}
  \city{Toronto}
  \country{Canada}
}
\email{aliredaq@yorku.ca}

\author{Hung Viet Pham}
\affiliation{%
  \institution{York University}
  \city{Toronto}
  \country{Canada}
}
\email{hvpham@yorku.ca}

\date{June 2024}

\begin{document}
\begin{abstract}

Large Language Models (LLMs) have demonstrated impressive performance in software engineering tasks. However, improving their accuracy in generating correct and reliable code remains challenging. Numerous prompt engineering techniques (PETs) have been developed to address this, but no single approach is universally optimal. Selecting the right PET for each query is difficult for two primary reasons: (1) interactive prompting techniques may not consistently deliver the expected benefits, especially for simpler queries, and (2) current automated prompt engineering methods lack adaptability and fail to fully utilize multi-stage responses.


To overcome these challenges, we propose \tool, a PET-agnostic selection model that uses code complexity as a proxy to classify queries and select the most appropriate PET. By incorporating contrastive learning, \tool effectively distinguishes between simple and complex problems, allowing it to choose PETs that are best suited for each query's complexity level.


Our evaluations on the MBPP and HumanEval benchmarks using GPT-3.5 Turbo and GPT-4o show up to a 1.9\% improvement in pass@1 accuracy, along with a 74.8\% reduction in token usage. Additionally, we provide both quantitative and qualitative results to demonstrate how \tool effectively selects the most appropriate techniques for each code generation query, further showcasing its efficiency in optimizing PET selection.
\end{abstract}

\ccsdesc[500]{Computing methodologies}
\ccsdesc[500]{Computing methodologies~Machine learning}
\ccsdesc[300]{Applied computing}
\ccsdesc[500]{Computing methodologies~Semantic networks}

\keywords{Prompt Engineering, Code Generation, Large Language Models}

\maketitle
\section{Introduction}
Recently Large Language Models (LLMs) have shown their promising performance in various software engineering tasks, such as unit test case generation~\cite{shin2024assessing,schafer2023empirical, shin2023domain}, automated bug repair~\cite{hossain2024deep, wei2023copiloting}, API specification~\cite{kim2024leveraging}.
Especially for code generation from natural language descriptions, LLMs demonstrate their impressive capability where code is generated with natural language descriptions~\cite{jiang2024survey, shin2021survey}.


Given the state-of-the-art LLMs are all closed-source, the most popular way to enhance the LLM's ability to generate accurate and reliable code is to utilize various prompt engineering techniques (PETs)~\cite{shin2023prompt,feng2024prompting}. For example, Some techniques ask LLMs to provide reasoning steps for solving problems~\cite{brown2020language, yao2024tree, yang2024buffer}, while others utilize LLMs to refine their output by prompting them to review and improve the code they generate~\cite{madaan2024self, chen2023teaching}.
In addition to these strategic PETs, some frameworks have been proposed that leverage LLMs~\cite{zhou2022large} or retrieve relevant instances from databases to automatically generate optimal prompts for questions~\cite{nashid2023retrieval}, a process known as automated prompt engineering.

Despite numerous studies focused on crafting the optimal prompt, not a single technique is optimally applicable to every query and the task of selecting a correct PET is not trivial. This is because of two key reasons: (1) interactive prompting techniques might be too costly and do not always provide the promised benefit especially when applied to simpler queries~\cite{huang2023large, chiang2024over}, and (2) existing automatic prompt engineering does not utilize the multiple stages of responses which is associated with the success of iterative PETs~\cite{zheng2023progressive, madaan2024self}. Not to mention, all of the auto prompt engineering techniques are not easily extended.

Prior work~\cite{zhao2023automatic} has proposed a framework to select the most appropriate PET for a given query based on feedback from LLMs. However, this approach focuses on reasoning tasks and requires implementation alongside language model execution, where the best answer is selected based on the outputs of various techniques. This makes it less practical and quite costly.

To provide a general low-cost solution to the PETs selection task, we propose \tool, a PET agnostic selection model that is not dependant on the pool of available PETs and can be easily adaptable and extendable to the ever-growing list of available advanced PETs. \tool integrates query complexity by using generated code complexity as a proxy using contrastive learning~\cite{khosla2020supervised}. 
Specifically, by incorporating generated code complexity, \tool can differentiate each query between simple and complex problems (i.e., requiring simple or complex code) which can help \tool select the appropriate PET that targets the relevant level of problem. Furthermore, we incorporate a wide range of PETs representing various categories~\cite{tony2024prompting}
including those PETs that have multi-round interactions with language models.

We evaluate \tool on two popular code generation benchmark datasets, MBPP and HumanEval. To ensure a fair evaluation, we apply 5-fold cross-validation with 80\% training and 20\% testing sets.
Our evaluation on GPT-3.5 Turbo and GPT-4o shows that \tool achieves an improvement of up to 1.9\%
in terms of pass@1 accuracy when compared with an individual PET while using as little as 74.8\%
fewer tokens on the HumanEval with GPT-4o. Our quantitative and qualitative results also demonstrate that \tool effectively selects appropriate techniques for each code generation query.
This paper makes the following contributions: 
\begin{itemize}
    \item \tool, a novel approach that automatically selects the most optimal prompting engineering techniques for each code generation query.
    \item An evaluation of \tool on two widely used benchmark datasets using two state-of-the-art LLMs.
    \item Quantitative and qualitative analyses provide insights into how \tool selects the appropriate PET.
\end{itemize}


\section{Background} \label{background}

\subsection{Automated Prompt Engineering}
Since Large Language Models (LLMs) are too large to fine-tune for every downstream task, prompt engineering has become a common approach to optimize performance across various tasks, including unseen ones. However, designing effective prompts for each task is a challenging process. Several studies have suggested reliable methods to improve language model performance, such as Chain-of-Thought and Self-correction prompting. Despite this, the question remains whether we can develop a system that automatically generates appropriate prompts for different queries. Previous studies~\cite{zhou2022large, do2024automatic} proposed frameworks for automatic instruction generation and selection, where several candidate prompts are generated by LLMs, and the best prompt is chosen from these candidates. Another approach involves retrieving similar queries from a database and using them to create a more effective prompt~\cite{nashid2023retrieval}. However, these automatic prompt engineering methods primarily focus on crafting a single optimal prompt for a given problem. There is limited research on how to design multi-round prompting, where multiple interactions with language models are used to refine the response. Crafting prompts based on the model's responses is crucial, as many state-of-the-art prompting techniques rely on self-generated answers to achieve optimal performance. Whether used for correction or evaluation, iterative interactions with language models play a key role in helping them generate better responses.

Technically, prompting technique selection is also a form of automatic prompt engineering, as it involves choosing the relatively appropriate prompt automatically. Unlike previous approaches, prompting technique selection considers whether the prompt should be crafted for a single or multiple iterations, allowing for multiple rounds of interaction. A previous study~\cite{zhao2023automatic} selects prompting techniques after each execution, which is costly and impractical in real-world applications, particularly when multiple techniques are considered as candidates. \tool is the first framework to select prompting techniques prior to execution. It employs a traditional deep learning model with contrastive learning to select the most suitable technique for each question, making it applicable and affordable even without the need to run language models.

\subsection{Prompt Engineering Challenges}
With the increasing number of prompting techniques being proposed and achieving state-of-the-art results on various benchmark datasets, a question arises: ``Can we apply the most advanced prompting techniques to every question?'' Unfortunately, the answer may be no. The first and most obvious issue is that using these advanced prompting techniques for every question is costly, as they often require multiple interactions with language models or involve crafting lengthy prompts with numerous examples. The second, and less well-known issue is that applying advanced prompting techniques to simpler questions can sometimes lead to incorrect answers. A recent study~\cite{chiang2024over} experimented on a variant of GSM8K, where all the answers to the questions in the dataset were explicitly stated in the questions themselves and could be obtained without any calculations. Surprisingly, the accuracy improves when language models are restricted from performing any calculations or reasoning steps, compared to when no instructions are specified. This suggests that unnecessary calculations and over-reasoning can lead to incorrect answers. There is another study~\cite{huang2023large} suggests that language models are not able to self-correct themselves. Self-correction is defined as a scenario where the model attempts to correct its initial responses purely based on its capabilities, without relying on external feedback. Many advanced prompting techniques leverage the self-correction ability of language models, such as Progressive Hint~\cite{zheng2023progressive} and Self-refine~\cite{madaan2024self}. However, research has shown that accuracy decreases with each iterative round. This suggests that the model struggles to identify and correct the specific incorrect parts. When the initial answer is correct, the model often changes the correct portion to something incorrect, resulting in a wrong answer. 

\tool learns to determine whether a question is easy or difficult by predicting the code complexity of the ground-truth code. This allows \tool to choose the relatively appropriate prompting techniques for each query, applying simpler techniques to easy problems and more advanced ones to difficult problems. This approach helps prevent over-reasoning and redundant calculations for easy questions, while also avoiding situations where the model changes a correct answer to an incorrect one.

\section{Approach} \label{approach}
\begin{figure}
    \centering
    \includegraphics[width=0.9\linewidth]{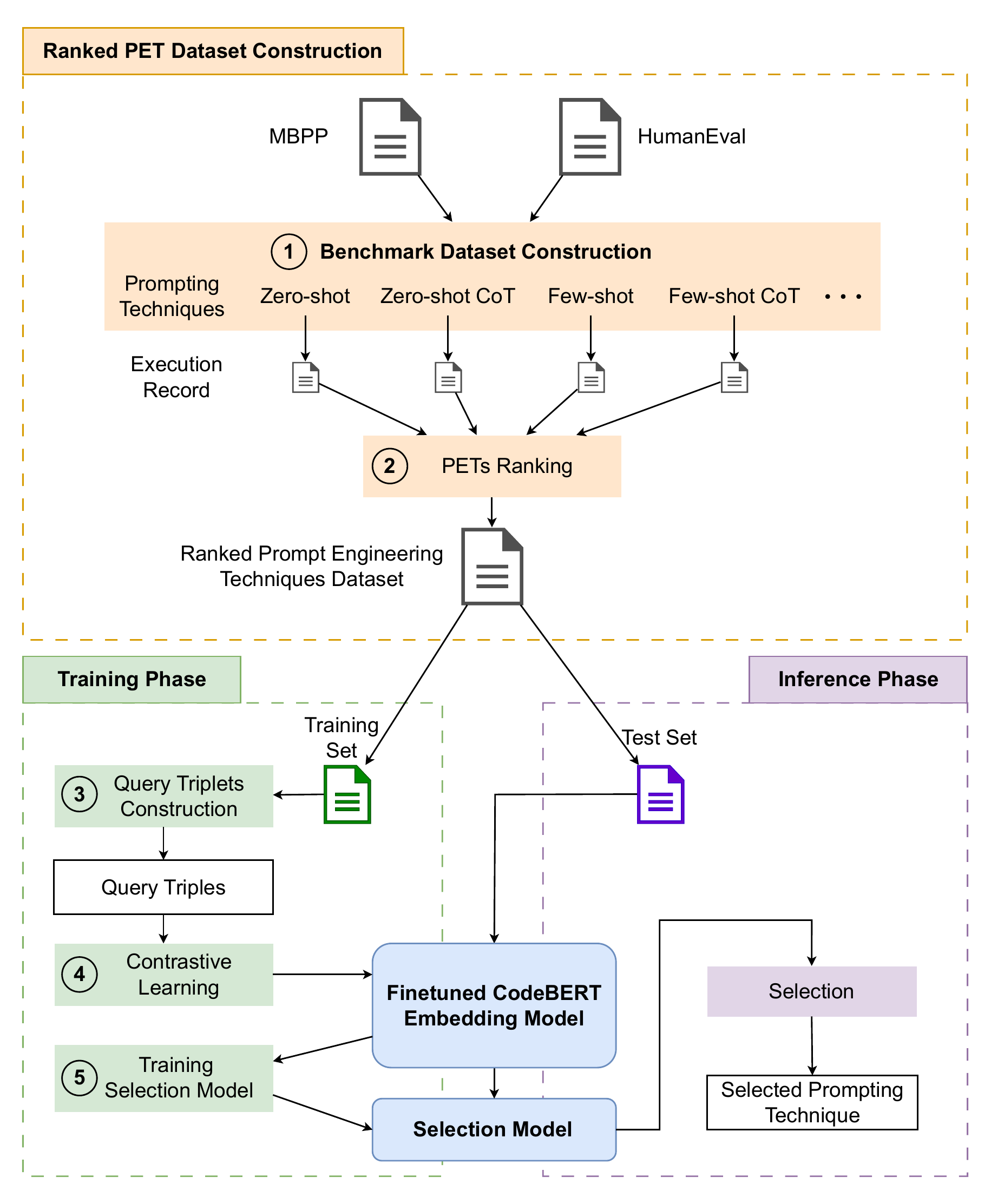}
    \vspace{-7pt}
    \caption{\tool Pipeline. 
    }
    \label{fig:pipeline}
    \vspace{-2pt}
\end{figure}

In this work, we propose \tool, a novel method to select suitable prompt engineering techniques (PETs) for each query. Figure~\ref{fig:pipeline} provides the overview of \tool. \tool is a supervised learning approach and since no such record of execution is available for various prompt engineering techniques we start off by building the data in the Dataset Construction phase (Section \ref{ground-truth}). \tool's model consists of two main parts: the embedding layer (Section \ref{sentence-embedding}) and the classification layer (Section \ref{selection}). Finally, we conduct a n-fold cross-validation evaluation to ensure that \tool is correctly evaluated.


\subsection{Ranked PET Dataset Construction} \label{ground-truth}

\tool is designed to be prompt engineering technique (PET) agnostic, we decide that unsupervised learning is the most appropriate approach. To train \tool, we first need to conduct a study to collect the dataset of execution records of various representative PETs such as Zero-shot, and Few-shot and rank each PET given their performance and cost for each query. Since numerous PETs could be employed for the code generation task, we select the most representative ones by choosing at least one technique from each fundamental strategic design category, such as root techniques, refinement-based techniques, and others, as defined in a recent study~\cite{tony2024prompting}. Detailed descriptions and implementations of these prompting techniques are provided in Section~\ref{prompting}.

\subsubsection{Benchmark Dataset Construction}
We choose the two most popular code generation datasets MBPP and HumanEval and benchmark the selected PETs on ChatGPT 3.5 and 4o (Step \circled{1}).
The responses were recorded along with the cost of the query in terms of the number of input and output tokens. Along with the query cost, the generated code complexity measured by five metrics is also recorded, the weighted sum of which is used as the overall complexity score.
Details of the metrics used are provided in Section~\ref{code_metrics}.



\subsubsection{PETs Ranking}
Once every technique has been benchmarked, we select the most appropriate one for each query with the highest $R\_Score_i$ (Step \circled{2}). Where $R\_Score_i$ for technique $i$ is calculated as:
\[ R\_Score_{i} = log(max^N_{j=1}(T\_tokens_j)) \times pass_i - log(T\_tokens_i) \]
Here, $T\_tokens$ is the sum of the number of input and output tokens required by PET $i$, and the $max(Max^N_{j=1}(T\_tokens_j))$ represents the highest number of required tokens across all prompting techniques for that query. The binary number $pass_i$ is 1 (i.e., the generated code passes all test cases) and 2 (i.e., at least one test case failed). Specifically, for techniques that fail to generate test passing code, the formula ensures that the score will be negative, and for successful techniques, the score will be positive. In all cases, the score is always inversely proportional to the number of required tokens. In the end, the technique that generates the correct code while requiring the fewest number of tokens will have the highest score. Since no technique uses the same number of tokens, there are no tied scores between the PETs, we can always choose the most appropriate one for each query. 



After this stage, we will obtain the Ranked PETs Dataset in which each entry includes the query string, the generated code, the number of tokens used, the complexity measures, and the most successful PET with the highest R\_score as the label.


\subsection{Fine-tuning CodeBERT Embedding Model} \label{sentence-embedding}
Based on their design, some PETs are better with more complex queries than others~\cite{khot2022decomposed}.
Given this finding, we want to incorporate the generated code complexity into our model decision-making to achieve the best prediction result. We accomplish this by tuning CodeBERT~\cite{feng2020codebert} embedding model utilizing conservative learning~\cite{khosla2020supervised}.
Specifically, the tuning process reshapes the embedding space so that queries with similar generated code complexity will be closer while dissimilar queries are placed farther apart.


\subsubsection{Query Triplets Construction} \label{query_triplets}
Contrastive learning performs optimization on query triplets each including an anchor query, a positive query, and a negative query~\cite{hoffer2015deep, wei2022clear}. Specifically, anchor queries are the original natural language questions, positive queries are either semantically equivalent to or share the same answer as the anchor queries~\cite{khosla2020supervised}, while negative queries are unrelated to both the anchor and positive queries. 

In this work, for a given query (i.e., the anchor query), we select positive queries as those with similar generated code complexity and negative queries as those with differing code complexity (Step \circled{3}).
For instance, given an anchor query ``\textit{Write a function to get the word with most number of occurrences in the given strings list.}'' with the generated code complexity score of 17, the positive query could be ``\textit{Write a python function to remove even numbers from a given list.}'' with the same code complexity score of 17, and the negative query could be ``\textit{Write a function to find the maximum product subarray of the given array.}'' with a much higher code complexity score of 58.

However, some queries may not have queries with the same code complexity score, we instead divided the entire training set into two categories: an easy set and a hard set. Queries with a code complexity lower than a specified threshold are placed in the easy set, while those exceeding the threshold are assigned to the hard set. We then randomly select a query from the same set as the anchor query to serve as its positive query. Conversely, a query is randomly selected from the opposite set to serve as the anchor query's negative query. To determine the optimal threshold for classifying the easy and hard sets, we conduct a grid search within the code complexity score range of 25 to 45, where more than 70\% of the scores are concentrated.
The configuration that yields the best result is selected as the optimal setting for the model.

\begin{figure}
    \centering
    \includegraphics[width=0.9\linewidth]{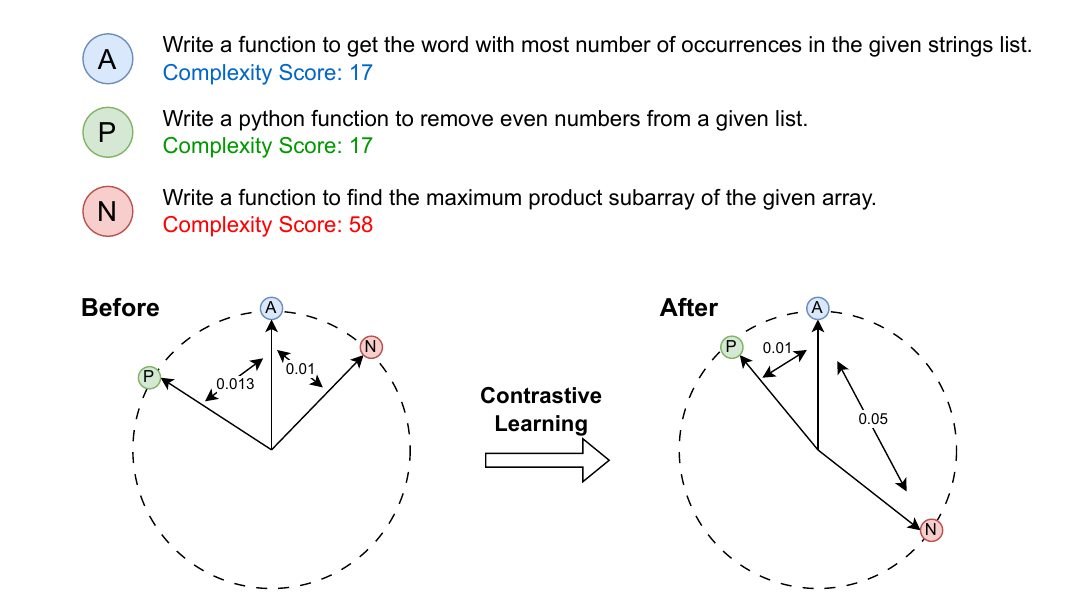}
    \vspace{-7pt}
    \caption{An example that demonstrates contrastive learning on a single anchor query.
    }
    \label{fig:contrastive}
    \vspace{-2pt}
\end{figure}
\subsubsection{Contrastive Learning} \label{contrastive}
Once the query triplets are constructed, we use them to fine-tune the CodeBERT sentence embedding model (Step \circled{4}). The objective of contrastive learning is to bring queries with similar features and complexity closer together while pushing unrelated queries with differing complexity further apart~\cite{oord2018representation}. When constructing the query triplets, we designate an input query as the anchor, treating queries with similar code complexity scores as positive examples, while those with dissimilar scores are used as negative examples. This design allows the model to learn semantic representations by associating anchor queries with their positive counterparts, positioning them closer within the embedding vector space. Conversely, we expect the model to push unrelated queries further apart from the anchor queries. Figure~\ref{fig:contrastive} illustrates the examples discussed in Section~\ref{query_triplets} to show the progress of contrastive learning in \tool. Before contrastive learning, the cosine distance between the anchor sentence (blue point) and the positive sentence (green point) is 0.013, which is greater than the distance between the anchor sentence and the negative sentence (red point), measured at 0.01. However, after contrastive learning, the positive sentence is brought closer to the anchor sentence in the embedding vector space, reducing the distance to 0.01, while the negative sentence is pushed farther away, increasing the distance to 0.05.

\tool's model architecture is built with the Sentence Transformer framework, specifically leveraging CodeBERT as a Transformer-based model for sentence embedding. First, the pre-trained CodeBERT model is used to extract embeddings for each word in sentences (in the anchor, positive, and negative queries). These word embeddings are aggregated with a pooling layer to create a fixed-size sentence-level embedding. The embedding model is fine-tuned by minimizing a Triplet Loss, which is computed based on the distances between the anchor-positive and anchor-negative query pairs:
\[ L = max(0, Distance_{anchor, positive}-Distance_{anchor, negative}+margin) \]
In short, the loss function is to learn an embedding space where semantically similar sentences are clustered together (small distance), and dissimilar sentences are far apart (large distance)~\cite{bredin2017tristounet}.
The $margin$ is a positive value (by default set to 1 in the model) that defines a minimum gap between the anchor-positive and anchor-negative distances. It ensures that the negative sentence is not simply pushed just outside the positive one but is kept at a meaningful distance. The $max$ function ensures the loss is non-negative, meaning if the distance between the negative and the anchor is already sufficiently large, the loss will be zero (i.e., no update is needed for this triplet). 


\subsection{Training Selection Model} \label{selection}
Once the embedding is computed, it can be used to extract a sentence embedding for any given query. The embedding will be used as input to \tool's three fully connected layers of neural network with ReLU activation function (Step \circled{5}). These layers are tasked with multi-class classification (i.e., PET selection). Specifically, the predicted technique is selected based on the highest probability according to the softmax function. These layers are trained normally using cross-entropy loss. It is important to note that the data used for training both the sentence embedding model and the selection model is within the training dataset and the model never sees the test set which is set aside to evaluate the model. For evaluation, we also record the probability of each class to calculate the MRR and nDCG metrics (described in Section~\ref{result}) for the results (Step \circled{6}).


\section{Experimental Setup} \label{setup}
In this section, we introduce the setup that we used to conduct our experiments. We first introduce the prompting techniques that are included in \tool selection pool, we then discuss the code complexity metrics, and finally, the experimental setting including the code generation datasets and the evaluation metrics.


\begin{table}
    \centering
    \caption{The prompting techniques used in the experiments. The `Strategic Category' column indicates the primary strategy of each technique, chosen from one of the five categories defined in the previous study~\cite{tony2024prompting}. The `Iteration' column specifies whether the technique requires multiple rounds of interaction with LLMs. The `Examples' column shows whether examples are included in the prompt construction. Lastly, the `Template' column outlines the specific prompt template used in the experiments.
    }
    \label{tab:techniques}
    \resizebox{\linewidth}{!}{
    \begin{tabular}{llccp{11cm}}
    \toprule
         &  Strategic Category&  Iteration&  Examples &Template\\
    \midrule
         Zero-shot~\cite{brown2020language}&  Root&  Single& \ding{55}   &Only generate the Python code for the following task. \textbf{\{Coding Task\}}.\\
    \midrule
         \multirow{3}{*}{Few-shot~\cite{brown2020language}}& \multirow{3}{*}{Root}& \multirow{3}{*}{Single}& \multirow{3}{*}{\ding{51}} &Here are some examples of how to generate the code. \newline \textbf{\{Three examples\}}. \newline How about this task? \textbf{\{Coding Task\}}.\\
    \midrule
         \multirow{2}{*}{Zero-shot CoT~\cite{kojima2022large}}&  \multirow{2}{*}{Reasoning}&  \multirow{2}{*}{Single}& \multirow{2}{*}{\ding{55}}  &Only generate the Python code for the following task. \textbf{\{Coding Task\}}. \newline Let's generate the code step by step.\\
    \midrule
         \multirow{3}{*}{Few-shot CoT~\cite{wei2022chain}}& \multirow{3}{*}{Reasoning}& \multirow{3}{*}{Single}&\multirow{3}{*}{\ding{51}}  &Here are some examples of how to generate the code step by step. \newline \textbf{\{Three examples with reasoning steps\}}. \newline How about this task? \textbf{\{Coding Task\}}.\\
    \midrule
         \multirow{2}{*}{Persona~\cite{white2023prompt}}& \multirow{2}{*}{Priming}& \multirow{2}{*}{Single}&\multirow{2}{*}{\ding{55}}  &You are a programming expert, especially good at Python. \newline Please complete the following task in Python: \textbf{\{Coding Task\}}.\\
    \midrule
         \multirow{7}{*}{Self-planning~\cite{jiang2023self}} &\multirow{7}{*}{Decomposition} &\multirow{7}{*}{Multiple} &\multirow{7}{*}{\ding{51}} &Plan Stage: \newline \textbf{\{Three examples of showing the Intent and Plan\}} \newline How about this intent: \textbf{\{Coding Task\}}. \\ \cmidrule{5-5}&&&&Implementation Stage: \newline \textbf{\{Coding Task\}}. \newline Please complete the task with the following plan in Python. \newline \textbf{\{Plan generated by the Plan Stage\}}.\\
    \midrule 
         \multirow{9}{*}{Self-refine~\cite{madaan2024self}}& \multirow{9}{*}{Refinement}& \multirow{9}{*}{Multiple}&\multirow{9}{*}{\ding{55}}& Initial Stage: \newline Only generate the Python code for the following task. \textbf{\{Coding Task\}}\\ \cmidrule{5-5}&&&&Reflection Stage: \newline Here is a code snippet: \textbf{\{Code generated by Initial Stage\}}. \newline Please review the code and suggest any improvements or identify any issues. \\ \cmidrule{5-5} &&&&Refinement Stage: \newline Here is a code snippet: \textbf{\{Code generated by Initial Stage\}}. \newline Based on the following feedback, refine the code: \newline \textbf{\{Feedback generated by Reflection Stage\}}. \\
    \midrule
         \multirow{5}{*}{Progressive Hint~\cite{zheng2023progressive}}& \multirow{5}{*}{Refinement}& \multirow{5}{*}{Multiple}&\multirow{5}{*}{\ding{55}}  &Initial Stage: \newline Please complete the following task in Python. \textbf{\{Coding Task\}}. \\ \cmidrule{5-5}&&&&Hint Stage: \newline Please complete the task in Python. \newline The answer is near to: \textbf{\{Code generated by Initial Stage\}}.\\
    \midrule
         \multirow{9}{*}{Self-debug~\cite{chen2023teaching}}& \multirow{9}{*}{Refinement}& \multirow{9}{*}{Multiple}&\multirow{9}{*}{\ding{55}}  &Initial Stage: \newline Only generate the Python code for the following task. \textbf{\{Coding Task\}} \newline Your code should pass the test: \textbf{\{One test case of the Coding Task\}}.\\ \cmidrule{5-5}&&&&Success Stage: \newline \textbf{\{Code generated by Initial Stage\}}. \newline Is the code above correct? If not, please fix it.\\ \cmidrule{5-5} &&&&Failure Stage: \newline \textbf{\{Code generated by Initial Stage\}}. \newline The code above is wrong. Please fix it.\\
    \bottomrule
    \end{tabular}
    }
    
\end{table}

\subsection{Prompt Engineering Techniques (PETs) for code generation} \label{prompting}
Table~\ref{tab:techniques} provides a summary of the PETs used in our experiment. To ensure a broad exploration of techniques, we selected at least one from each category as stated in the recent work~\cite{tony2024prompting}. These prompting techniques are classified into five categories based on their core concepts: root techniques, refinement-based techniques, decomposition-based techniques, reasoning-based techniques, and priming techniques. The ``Strategic Category'' column indicates the categorization of each prompting technique, while the ``Iteration'' column specifies whether the technique involves iterative interactions with the language models. The ``Examples'' column shows whether the technique includes examples in the prompt to guide the language models on how to answer the questions. The ``Template'' column demonstrates the prompting templates we used for each technique. For techniques with multiple iterations, we provided specific prompting templates for each stage. We briefly go through each PET and provide some pros and cons to emphasize that no one PET is optimal for all cases.

\textbf{Root PETs: Zero-shot and Few-shot} Root PETs directly query LLMs for answers. Zero-shot and Few-shot~\cite{brown2020language} are two examples of root PETs where Zero-shot provides no additional example and Few-shot includes several examples. While it is convenient and requires no domain-specific input, Zero-shot performance may be limited when the model encounters unfamiliar tasks. The added examples in Few-shot PET improve LLMs' ability to handle unseen tasks but are not trivial to craft~\cite{dang2022prompt, liu2021makes, perez2021true} and can negatively impact the performance if given incorrectly~\cite{rubin2021learning, lu2021fantastically}.






\textbf{Reasoning PETs: Zero-shot/Few-shot Chain-of-Thought (CoT)} are reasoning-based techniques that query LLMs to explain intermediate reasoning steps while generating answers~\cite{wei2022chain, kojima2022large}.
It enables LLMs to produce more coherent and accurate results.
The zero-shot and few-shot CoT differ in the presence of examples: zero-shot CoT does not include examples while few-shot CoT offers additional reasoning examples in the query.
Despite the performance improvements similar limitations persist: 
zero-shot CoT can yield unreliable results on unfamiliar tasks, and the need for carefully crafted prompts with examples remains a challenge with few-shot CoT.

\textbf{Priming PETs: Persona} is a PET that 
LLM is guided to take on a specific identity or personality
based on expertise, tone, or role. 
This ``persona'' helps make the communication with LLMs consistent, but a too specific persona can lead to restrictive communication.

\textbf{Decomposition PETs: Self-planning} involves having the LLMs create a mental blueprint or set of steps before answering a question. This is particularly useful for complex tasks that require a structured approach (e.g., solving math problems)~\cite{zhou2022least}. On the one hand, this can provide structure to the solution but on the other, if the initial plan is incorrect, the entire response may be off track.

\textbf{Refinement PETs: Self-refine, Progressive Hint, and Self-debug} take a different approach by having the LLM interact with its own response after generating it. Specifically, Self-refine~\cite{madaan2024self}, Progressive Hint~\cite{zheng2023progressive}, and Self-debug~\cite{chen2023teaching} ask the LLM to review its answers, use its answers as hints, and correct its output based on the execution result of test cases. While self-refine can sometimes correct itself, the errors might still pass notice. Progressive Hint also suffers from similar pitfalls where the first hint can be incorrect and create a domino effect. Finally, with the help of the external test cases, self-debug can sometimes correct itself, however, the debugging process is not perfect and sometimes LLM can over-correct itself thus generating the wrong answer.

\subsection{Code complexity metrics} \label{code_metrics}
\tool utilize 
five popular code complexity metrics: Line of Code, Cyclomatic Complexity, Halstead Complexity, Cognitive Complexity, and Maintainability Index~\cite{zhang2007predicting, jiang2008comparing, shin2008empirical}
to aid with the contrastive learning step:
\textbf{Line Complexity} is also known as Lines of Code (LOC), which measures the number of lines in a codebase. 
In this study, Line Complexity is calculated using Physical Lines of Code (PLOC), which excludes comment lines and focuses solely on the program's source code.
\textbf{Cyclomatic Complexity~\cite{ebert2016cyclomatic}}
counts the number of independent paths through the code.
Higher cyclomatic complexity indicates more potential paths, increasing the testing effort and potentially reducing maintainability. 
\textbf{Halstead Complexity~\cite{hariprasad2017software}} evaluates code complexity from both linguistic and mathematical perspectives, based on the number of operators and operands. 
\textbf{Cognitive Complexity~\cite{campbell2018cognitive}} measures how difficult code is for a human to understand by considering factors like nesting depth and control structures such as if, switch, and for loops. Unlike cyclomatic complexity, it focuses on readability and the mental effort required to follow the code.
\textbf{Maintainability Index~\cite{welker2001software}} is a composite metric that predicts the ease of maintaining a software system, combining factors like cyclomatic complexity, Halstead complexity, and lines of code. It ranges from 0 (difficult to maintain) to 100 (easy to maintain), with higher values indicating better maintainability. 
In this study, custom code was used to calculate LOC, the Radon package was used to calculate Cyclomatic Complexity, Halstead Complexity, and Maintainability Index, and Cognitive Complexity was computed with the cognitive-complexity Python package.

\subsection{Experiment Settings}
\noindent{\textit{\textbf{Benchmark datasets}}}
We used two of the most widely used code generation benchmark datasets to train the model and evaluate \tool's performance: HumanEval~\cite{chen2021evaluating} and MBPP~\cite{austin2021program}. Both datasets provide test cases so that generated code can be functionally evaluated and the pass@k metric can be calculated for evaluation.


\noindent{\textit{\textbf{Ranking Evaluation Metrics}}
Since \tool ranks all the prompting techniques based on the probability of softmax layer output, we applied two popular metrics, Mean Reciprocal Rank (MRR)~\cite{voorhees1999trec,radev2002evaluating} and Normalized Discounted Cumulative Gain (nDCG)~\cite{jarvelin2002cumulated}, to evaluate \tool. These metrics are used extensively in the domain of information retrieval and they measure the ability of a system in recommendation tasks. The Mean Reciprocal Rank measures the effectiveness of a system in returning relevant results or answers by focusing on the rank of the first correct. On the other hand, Normalized Discounted Cumulative Gain (nDCG) measures the quality of ranked results based on the relevance of each result and the position in which they appear in a ranking list. With these two metrics, we can thoroughly evaluate \tool's ability to recommend and rank the appropriate PET.

\noindent{\textit{\textbf{Environmental Settings}}
We utilize a machine with an 8-core processor AMD Ryzen 7 pro 5845 and an NVIDIA RTX3060 to train \tool.
To better evaluate \tool we applied 5-fold cross-validation with 80-20 train-test split. Note that the sentence embedding model and the selection model were only trained on the training set to prevent test data leakage into the sentence embedding model, which could otherwise impact the performance of the selection model.
We fine-tune the sentence embedding model for fifteen epochs and select the model with the best performance (the highest value of Cosine Accuracy) on the validation set to train the selection model. For the selection model, we train it for 10 epochs and select the model with the best performance (the highest value of nDCG) on the validation set to choose prompting techniques for each instance in the test set.



\section{Result} \label{result}
In this section, we evaluate \tool and present the findings when exploring three research questions. RQ1 explores how various PETs perform on different types of code generation with different complexity (Section ~\ref{RQ1}). In RQ2, we compare \tool performance against other baselines on two code generation benchmarks using two versions of GPT (Section~\ref{RQ2}). Finally, we analyze \tool's performance in quantitative and qualitative analysis (Section~\ref{RQ3}).

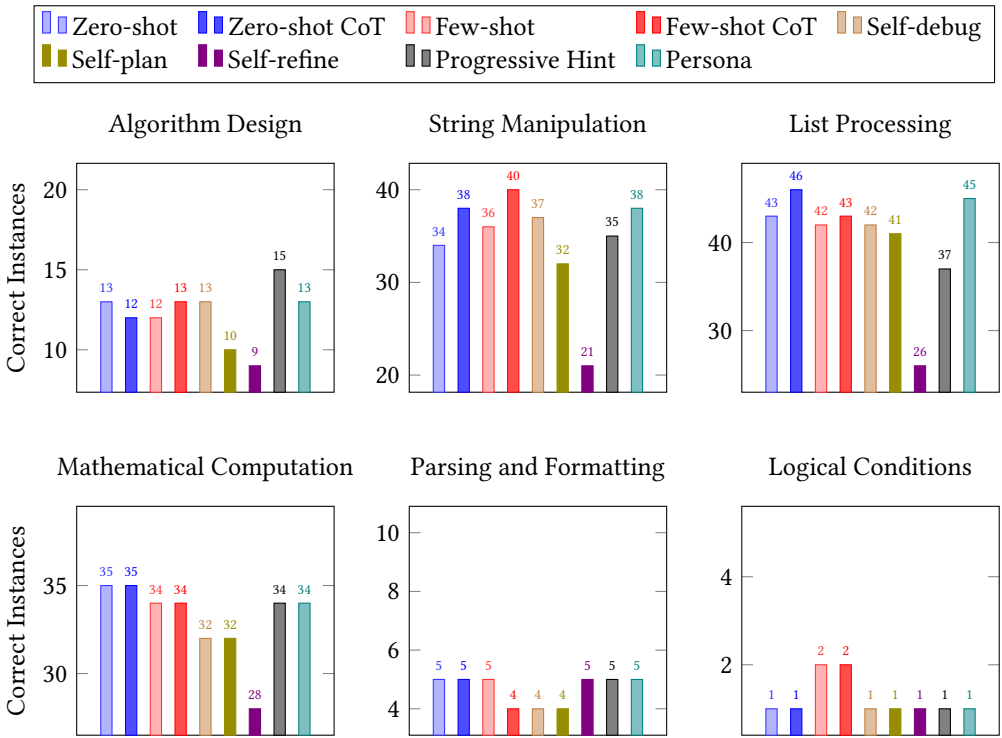
\begin{figure}[t!]
\pgfplotsset{compat=1.18}

\begin{tikzpicture}
\begin{groupplot}[
    group style={
        group name=my plots, 
        group size=3 by 2, 
        xlabels at=edge bottom, 
        ylabels at=edge left,
        vertical sep = 1.5cm
    },
    height=0.33\linewidth, 
    width=0.36\linewidth, 
    ylabel={Correct Instances},
    ybar, 
    enlargelimits=0.15, 
    symbolic x coords={A, B, C, D, E, F, G, H, I}, 
    xtick=\empty,
    legend style={
        at={(1.7,1.35)}, 
        anchor=south, 
        legend columns=5, 
        /tikz/every even column/.append style={column sep=0.2cm},
        legend cell align={left}
    },
]

\nextgroupplot[
    title={Algorithm Design},
    bar width = 0.15cm,
    nodes near coords,
    nodes near coords style={
        font=\fontsize{5pt}{4pt}\selectfont 
    },
    ymax=20,
]
\addplot+[bar shift=0cm, fill=blue!30, draw=blue!80, text=blue!80] coordinates {(A,13)}; \addlegendentry{Zero-shot}
\addplot+[bar shift=0cm, fill=blue!70, draw=blue!100, text=blue!100] coordinates {(B,12)}; \addlegendentry{Zero-shot CoT}
\addplot+[bar shift=0cm, fill=red!30, draw=red!80, text=red!80] coordinates {(C,12)}; \addlegendentry{Few-shot}
\addplot+[bar shift=0cm, fill=red!70, draw=red!100, text=red!100] coordinates {(D,13)}; \addlegendentry{Few-shot CoT}
\addplot+[bar shift=0cm, fill=brown!50, draw=brown!100, text=brown!100] coordinates {(E,13)}; \addlegendentry{Self-debug}
\addplot+[bar shift=0cm, fill=olive, draw=olive, text=olive] coordinates {(F,10)}; \addlegendentry{Self-plan}
\addplot+[bar shift=0cm, fill=violet, draw=violet, text=violet] coordinates {(G,9)}; \addlegendentry{Self-refine}
\addplot+[bar shift=0cm, fill=black!50, draw=black!100, text=black!100] coordinates {(H,15)}; \addlegendentry{Progressive Hint}
\addplot+[bar shift=0cm, fill=teal!50, draw=teal!100, text=teal!100] coordinates {(I,13)}; \addlegendentry{Persona}

\nextgroupplot[
    title={String Manipulation},
    bar width = 0.15cm,
    nodes near coords,
    nodes near coords style={
        font=\fontsize{5pt}{4pt}\selectfont 
    },
    ymin=21,
    ymax=40,
]
\addplot+[bar shift=0cm, fill=blue!30, draw=blue!80, text=blue!80] coordinates {(A,34)}; 
\addplot+[bar shift=0cm, fill=blue!70, draw=blue!100, text=blue!100] coordinates {(B,38)}; 
\addplot+[bar shift=0cm, fill=red!30, draw=red!80, text=red!80] coordinates {(C,36)}; 
\addplot+[bar shift=0cm, fill=red!70, draw=red!100, text=red!100] coordinates {(D,40)}; 
\addplot+[bar shift=0cm, fill=brown!50, draw=brown!100, text=brown!100] coordinates {(E,37)}; 
\addplot+[bar shift=0cm, fill=olive, draw=olive, text=olive] coordinates {(F,32)}; 
\addplot+[bar shift=0cm, fill=violet, draw=violet, text=violet] coordinates {(G,21)}; 
\addplot+[bar shift=0cm, fill=black!50, draw=black!100, text=black!100] coordinates {(H,35)}; 
\addplot+[bar shift=0cm, fill=teal!50, draw=teal!100, text=teal!100] coordinates {(I,38)}; 

\nextgroupplot[
    title={List Processing},
    bar width = 0.15cm,
    nodes near coords,
    nodes near coords style={
        font=\fontsize{5pt}{4pt}\selectfont 
    },
    ymin=26,
    ymax=46,
]
\addplot+[bar shift=0cm, fill=blue!30, draw=blue!80, text=blue!80] coordinates {(A,43)}; 
\addplot+[bar shift=0cm, fill=blue!70, draw=blue!100, text=blue!100] coordinates {(B,46)}; 
\addplot+[bar shift=0cm, fill=red!30, draw=red!80, text=red!80] coordinates {(C,42)}; 
\addplot+[bar shift=0cm, fill=red!70, draw=red!100, text=red!100] coordinates {(D,43)}; 
\addplot+[bar shift=0cm, fill=brown!50, draw=brown!100, text=brown!100] coordinates {(E,42)}; 
\addplot+[bar shift=0cm, fill=olive, draw=olive, text=olive] coordinates {(F,41)}; 
\addplot+[bar shift=0cm, fill=violet, draw=violet, text=violet] coordinates {(G,26)}; 
\addplot+[bar shift=0cm, fill=black!50, draw=black!100, text=black!100] coordinates {(H,37)}; 
\addplot+[bar shift=0cm, fill=teal!50, draw=teal!100, text=teal!100] coordinates {(I,45)}; 

\nextgroupplot[
    title={Mathematical Computation},
    bar width = 0.15cm,
    nodes near coords,
    nodes near coords style={
        font=\fontsize{5pt}{4pt}\selectfont 
    },
    ymin=28,
    ymax=38,
]
\addplot+[bar shift=0cm, fill=blue!30, draw=blue!80, text=blue!80] coordinates {(A,35)}; 
\addplot+[bar shift=0cm, fill=blue!70, draw=blue!100, text=blue!100] coordinates {(B,35)}; 
\addplot+[bar shift=0cm, fill=red!30, draw=red!80, text=red!80] coordinates {(C,34)}; 
\addplot+[bar shift=0cm, fill=red!70, draw=red!100, text=red!100] coordinates {(D,34)}; 
\addplot+[bar shift=0cm, fill=brown!50, draw=brown!100, text=brown!100] coordinates {(E,32)}; 
\addplot+[bar shift=0cm, fill=olive, draw=olive, text=olive] coordinates {(F,32)}; 
\addplot+[bar shift=0cm, fill=violet, draw=violet, text=violet] coordinates {(G,28)}; 
\addplot+[bar shift=0cm, fill=black!50, draw=black!100, text=black!100] coordinates {(H,34)}; 
\addplot+[bar shift=0cm, fill=teal!50, draw=teal!100, text=teal!100] coordinates {(I,34)}; 

\nextgroupplot[
    title={Parsing and Formatting},
    bar width = 0.15cm,
    nodes near coords,
    nodes near coords style={
        font=\fontsize{5pt}{4pt}\selectfont 
    },
    ymax=10,
]
\addplot+[bar shift=0cm, fill=blue!30, draw=blue!80, text=blue!80] coordinates {(A,5)}; 
\addplot+[bar shift=0cm, fill=blue!70, draw=blue!100, text=blue!100] coordinates {(B,5)}; 
\addplot+[bar shift=0cm, fill=red!30, draw=red!80, text=red!80] coordinates {(C,5)}; 
\addplot+[bar shift=0cm, fill=red!70, draw=red!100, text=red!100] coordinates {(D,4)}; 
\addplot+[bar shift=0cm, fill=brown!50, draw=brown!100, text=brown!100] coordinates {(E,4)}; 
\addplot+[bar shift=0cm, fill=olive, draw=olive, text=olive] coordinates {(F,4)}; 
\addplot+[bar shift=0cm, fill=violet, draw=violet, text=violet] coordinates {(G,5)}; 
\addplot+[bar shift=0cm, fill=black!50, draw=black!100, text=black!100] coordinates {(H,5)}; 
\addplot+[bar shift=0cm, fill=teal!50, draw=teal!100, text=teal!100] coordinates {(I,5)}; 

\nextgroupplot[
    title={Logical Conditions},
    bar width = 0.15cm,
    nodes near coords,
    nodes near coords style={
        font=\fontsize{5pt}{4pt}\selectfont 
    },
    ymax=5,
]
\addplot+[bar shift=0cm, fill=blue!30, draw=blue!80, text=blue!80] coordinates {(A,1)}; 
\addplot+[bar shift=0cm, fill=blue!70, draw=blue!100, text=blue!100] coordinates {(B,1)}; 
\addplot+[bar shift=0cm, fill=red!30, draw=red!80, text=red!80] coordinates {(C,2)}; 
\addplot+[bar shift=0cm, fill=red!70, draw=red!100, text=red!100] coordinates {(D,2)}; 
\addplot+[bar shift=0cm, fill=brown!50, draw=brown!100, text=brown!100] coordinates {(E,1)}; 
\addplot+[bar shift=0cm, fill=olive, draw=olive, text=olive] coordinates {(F,1)}; 
\addplot+[bar shift=0cm, fill=violet, draw=violet, text=violet] coordinates {(G,1)}; 
\addplot+[bar shift=0cm, fill=black!50, draw=black!100, text=black!100] coordinates {(H,1)}; 
\addplot+[bar shift=0cm, fill=teal!50, draw=teal!100, text=teal!100] coordinates {(I,1)};

\end{groupplot}

\end{tikzpicture}
\caption{The distribution of correct instances across nine PETs
on the HumanEval dataset using GPT-4o.}
\label{HumanEval_distribution}
\end{figure}
\subsection{RQ1. How do various PETs perform on different types of code generation with different complexity?} \label{RQ1}
In this research question, we aim to explore the relationship between the code generation types and code complexity to inform our design decisions to incorporate query embedding and generated code complexity in \tool.


\subsubsection{RQ1.1 Do different PETs excel at generating code for different types of tasks?}
To explore the first part of the question, 
we first manually categorize questions from the MBPP and HumanEval datasets into six different types of tasks for which the generated code is responsible: Algorithm Design, String Manipulation, List Processing, Mathematical Computation, Parsing and Formatting, and Logical Conditions.


Specifically, we applied the following definition to perform the labeling:
\begin{itemize}
    \item \textbf{Algorithm Design} involves writing code to solve problems using specific approaches or procedures. Algorithm design includes tasks like designing search algorithms (e.g., binary search), sorting (e.g., quicksort), and dynamic programming. The focus is on the logic and structure required to solve problems efficiently.
    \item \textbf{String Manipulation} deals with operations related to handling text data, such as modifying, concatenating, splitting, and searching within strings. Common tasks include pattern matching (using regular expressions), converting cases (e.g., uppercase to lowercase), and formatting strings for output. 
    \item \textbf{List Processing} involves handling collections or arrays of data. Operations include iterating through lists, filtering, mapping, sorting, and transforming data. Tasks like merging multiple lists or finding elements based on specific conditions also fall under this category.
    \item \textbf{Mathematical Computation} covers tasks that involve performing mathematical operations, such as arithmetic, algebra, trigonometry, or calculus. Examples include calculating averages, finding prime numbers, performing matrix operations, or solving equations.
    \item \textbf{Parsing} refers to interpreting structured data, such as converting a string into a number, extracting values from JSON or XML, or reading configuration files.
    \item \textbf{Formatting} involves preparing data for output, such as formatting dates, numbers, or aligning text for display.
    \item \textbf{Logical Conditions} involves decision-making in code, where you use conditions to control the flow of the program (e.g., if-else statements, switch cases). Logical conditions help programs execute different paths based on input or state, such as checking if a number is even or odd, or deciding which function to call based on user input.
\end{itemize}

Figure~\ref{HumanEval_distribution} presents the distribution of correct instances across different PETs on the HumanEval dataset using GPT-4o. For each task, some PETs are more effective than others. 
For example, Progressive Hint yields the highest number of correct instances in Algorithm Design, while for String Manipulation, the most successful technique shifts to Few-shot CoT. This finding suggests that each technique excels in different tasks, indicating its unique area of expertise. As a result, we included Category Selection as one of our baselines in RQ2 to explore whether choosing PETs directly based on the specific code generation tasks could help identify the most suitable technique for each question.

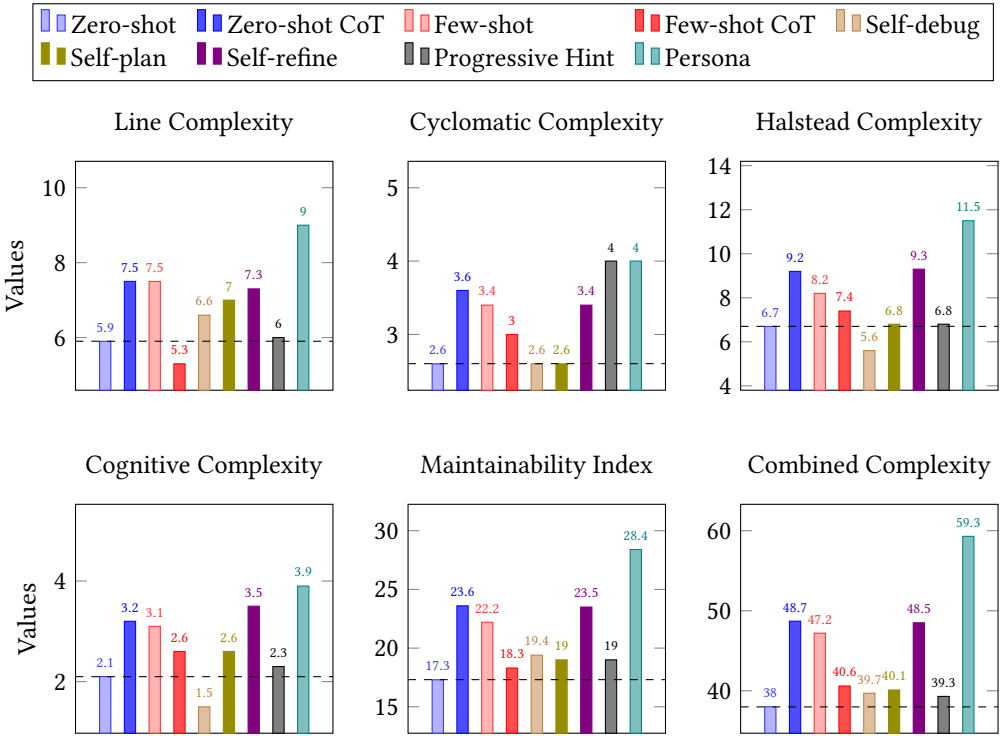
\begin{figure}[t!]
\pgfplotsset{compat=1.18}

\begin{tikzpicture}
\begin{groupplot}[
    group style={
        group name=my plots, 
        group size=3 by 2, 
        xlabels at=edge bottom, 
        ylabels at=edge left,
        vertical sep = 1.5cm
    },
    height=0.33\linewidth, 
    width=0.36\linewidth, 
    ylabel={Values},
    ybar, 
    enlargelimits=0.15, 
    symbolic x coords={A, B, C, D, E, F, G, H, I}, 
    xtick=\empty,
    legend style={
        at={(1.7,1.35)}, 
        anchor=south, 
        legend columns=5, 
        /tikz/every even column/.append style={column sep=0.2cm},
        legend cell align={left}
    },
]

\nextgroupplot[
    title={Line Complexity},
    bar width = 0.15cm,
    nodes near coords,
    nodes near coords style={
        font=\fontsize{5pt}{4pt}\selectfont 
    },
    ymax=10,
]
\addplot+[bar shift=0cm, fill=blue!30, draw=blue!80, text=blue!80] coordinates {(A,5.9)}; \addlegendentry{Zero-shot}
\addplot+[bar shift=0cm, fill=blue!70, draw=blue!100, text=blue!100] coordinates {(B,7.5)}; \addlegendentry{Zero-shot CoT}
\addplot+[bar shift=0cm, fill=red!30, draw=red!80, text=red!80] coordinates {(C,7.5)}; \addlegendentry{Few-shot}
\addplot+[bar shift=0cm, fill=red!70, draw=red!100, text=red!100] coordinates {(D,5.3)}; \addlegendentry{Few-shot CoT}
\addplot+[bar shift=0cm, fill=brown!50, draw=brown!100, text=brown!100] coordinates {(E,6.6)}; \addlegendentry{Self-debug}
\addplot+[bar shift=0cm, fill=olive, draw=olive, text=olive] coordinates {(F,7)}; \addlegendentry{Self-plan}
\addplot+[bar shift=0cm, fill=violet, draw=violet, text=violet] coordinates {(G,7.3)}; \addlegendentry{Self-refine}
\addplot+[bar shift=0cm, fill=black!50, draw=black!100, text=black!100] coordinates {(H,6)}; \addlegendentry{Progressive Hint}
\addplot+[bar shift=0cm, fill=teal!50, draw=teal!100, text=teal!100] coordinates {(I,9)}; \addlegendentry{Persona}
\draw[dashed] (rel axis cs:0,0.214) -- (rel axis cs:1,0.214); 

\nextgroupplot[
    title={Cyclomatic Complexity},
    bar width = 0.15cm,
    nodes near coords,
    nodes near coords style={
        font=\fontsize{5pt}{4pt}\selectfont 
    },
    ymax=5,
]
\addplot+[bar shift=0cm, fill=blue!30, draw=blue!80, text=blue!80] coordinates {(A,2.6)}; 
\addplot+[bar shift=0cm, fill=blue!70, draw=blue!100, text=blue!100] coordinates {(B,3.6)}; 
\addplot+[bar shift=0cm, fill=red!30, draw=red!80, text=red!80] coordinates {(C,3.4)}; 
\addplot+[bar shift=0cm, fill=red!70, draw=red!100, text=red!100] coordinates {(D,3)}; 
\addplot+[bar shift=0cm, fill=brown!50, draw=brown!100, text=brown!100] coordinates {(E,2.6)}; 
\addplot+[bar shift=0cm, fill=olive, draw=olive, text=olive] coordinates {(F,2.6)}; 
\addplot+[bar shift=0cm, fill=violet, draw=violet, text=violet] coordinates {(G,3.4)}; 
\addplot+[bar shift=0cm, fill=black!50, draw=black!100, text=black!100] coordinates {(H,4)}; 
\addplot+[bar shift=0cm, fill=teal!50, draw=teal!100, text=teal!100] coordinates {(I,4)}; 
\draw[dashed] (rel axis cs:0,0.116) -- (rel axis cs:1,0.116); 

\nextgroupplot[
    title={Halstead Complexity},
    bar width = 0.15cm,
    nodes near coords,
    nodes near coords style={
        font=\fontsize{5pt}{4pt}\selectfont 
    },
    ymin=5,
    ymax=13,
]
\addplot+[bar shift=0cm, fill=blue!30, draw=blue!80, text=blue!80] coordinates {(A,6.7)}; 
\addplot+[bar shift=0cm, fill=blue!70, draw=blue!100, text=blue!100] coordinates {(B,9.2)}; 
\addplot+[bar shift=0cm, fill=red!30, draw=red!80, text=red!80] coordinates {(C,8.2)}; 
\addplot+[bar shift=0cm, fill=red!70, draw=red!100, text=red!100] coordinates {(D,7.4)}; 
\addplot+[bar shift=0cm, fill=brown!50, draw=brown!100, text=brown!100] coordinates {(E,5.6)}; 
\addplot+[bar shift=0cm, fill=olive, draw=olive, text=olive] coordinates {(F,6.8)}; 
\addplot+[bar shift=0cm, fill=violet, draw=violet, text=violet] coordinates {(G,9.3)}; 
\addplot+[bar shift=0cm, fill=black!50, draw=black!100, text=black!100] coordinates {(H,6.8)}; 
\addplot+[bar shift=0cm, fill=teal!50, draw=teal!100, text=teal!100] coordinates {(I,11.5)}; 
\draw[dashed] (rel axis cs:0,0.279) -- (rel axis cs:1,0.279); 

\nextgroupplot[
    title={Cognitive Complexity},
    bar width = 0.15cm,
    nodes near coords,
    nodes near coords style={
        font=\fontsize{5pt}{4pt}\selectfont 
    },
    ymax=5,
]
\addplot+[bar shift=0cm, fill=blue!30, draw=blue!80, text=blue!80] coordinates {(A,2.1)}; 
\addplot+[bar shift=0cm, fill=blue!70, draw=blue!100, text=blue!100] coordinates {(B,3.2)}; 
\addplot+[bar shift=0cm, fill=red!30, draw=red!80, text=red!80] coordinates {(C,3.1)}; 
\addplot+[bar shift=0cm, fill=red!70, draw=red!100, text=red!100] coordinates {(D,2.6)}; 
\addplot+[bar shift=0cm, fill=brown!50, draw=brown!100, text=brown!100] coordinates {(E,1.5)}; 
\addplot+[bar shift=0cm, fill=olive, draw=olive, text=olive] coordinates {(F,2.6)}; 
\addplot+[bar shift=0cm, fill=violet, draw=violet, text=violet] coordinates {(G,3.5)}; 
\addplot+[bar shift=0cm, fill=black!50, draw=black!100, text=black!100] coordinates {(H,2.3)}; 
\addplot+[bar shift=0cm, fill=teal!50, draw=teal!100, text=teal!100] coordinates {(I,3.9)}; 
\draw[dashed] (rel axis cs:0,0.247) -- (rel axis cs:1,0.247); 

\nextgroupplot[
    title={Maintainability Index},
    bar width = 0.15cm,
    nodes near coords,
    nodes near coords style={
        font=\fontsize{5pt}{4pt}\selectfont 
    },
    ymin=15,
    ymax=30,
]
\addplot+[bar shift=0cm, fill=blue!30, draw=blue!80, text=blue!80] coordinates {(A,17.3)}; 
\addplot+[bar shift=0cm, fill=blue!70, draw=blue!100, text=blue!100] coordinates {(B,23.6)}; 
\addplot+[bar shift=0cm, fill=red!30, draw=red!80, text=red!80] coordinates {(C,22.2)}; 
\addplot+[bar shift=0cm, fill=red!70, draw=red!100, text=red!100] coordinates {(D,18.3)}; 
\addplot+[bar shift=0cm, fill=brown!50, draw=brown!100, text=brown!100] coordinates {(E,19.4)}; 
\addplot+[bar shift=0cm, fill=olive, draw=olive, text=olive] coordinates {(F,19.0)}; 
\addplot+[bar shift=0cm, fill=violet, draw=violet, text=violet] coordinates {(G,23.5)}; 
\addplot+[bar shift=0cm, fill=black!50, draw=black!100, text=black!100] coordinates {(H,19.0)}; 
\addplot+[bar shift=0cm, fill=teal!50, draw=teal!100, text=teal!100] coordinates {(I,28.4)}; 
\draw[dashed] (rel axis cs:0,0.234) -- (rel axis cs:1,0.234); 

\nextgroupplot[
    title={Combined Complexity},
    bar width = 0.15cm,
    nodes near coords,
    nodes near coords style={
        font=\fontsize{5pt}{4pt}\selectfont 
    },
    ymin=38,
    ymax=60,
]
\addplot+[bar shift=0cm, fill=blue!30, draw=blue!80, text=blue!80] coordinates {(A,38.0)}; 
\addplot+[bar shift=0cm, fill=blue!70, draw=blue!100, text=blue!100] coordinates {(B,48.7)}; 
\addplot+[bar shift=0cm, fill=red!30, draw=red!80, text=red!80] coordinates {(C,47.2)}; 
\addplot+[bar shift=0cm, fill=red!70, draw=red!100, text=red!100] coordinates {(D,40.6)}; 
\addplot+[bar shift=0cm, fill=brown!50, draw=brown!100, text=brown!100] coordinates {(E,39.7)}; 
\addplot+[bar shift=0cm, fill=olive, draw=olive, text=olive] coordinates {(F,40.1)}; 
\addplot+[bar shift=0cm, fill=violet, draw=violet, text=violet] coordinates {(G,48.5)}; 
\addplot+[bar shift=0cm, fill=black!50, draw=black!100, text=black!100] coordinates {(H,39.3)}; 
\addplot+[bar shift=0cm, fill=teal!50, draw=teal!100, text=teal!100] coordinates {(I,59.3)}; 
\draw[dashed] (rel axis cs:0,0.115) -- (rel axis cs:1,0.115);

\end{groupplot}
\end{tikzpicture}
\caption{The distribution of code complexity scores for the ground-truth code, correctly answered by each
PET
across six code complexity metrics on the MBPP dataset using GPT-3.5 Turbo.}
\label{MBPP_complexity}
\end{figure}

\subsubsection{RQ1.2 Do different PETs excel at generating code of different complexity?}
Apart from task types, we also explored if code complexity can inform the correct PET.
We hypothesized that simpler techniques might perform better on easier questions (i.e., requiring less complex code), while more complex techniques could be more effective on harder ones (i.e., requiring more complex code). To test this, we applied five code complexity metrics mentioned previously 
to the ground-truth code for each instance in the MBPP and HumanEval datasets. To account for multiple aspects of code complexity, we aggregate all the complexity scores into a single value called Combined Complexity, which serves as the final complexity score for each instance.

Figure~\ref{MBPP_complexity} demonstrates the distribution of code complexity scores for the ground-truth code, correctly answered by each PET, across six code complexity metrics on the MBPP dataset using GPT-3.5 Turbo. We can observe that the code complexity score of the ground truth solutions for questions that are answered correctly in zero-shot is lower than that of most techniques across all code complexity metrics. For example, in terms of line complexity, all PETs except Few-shot CoT achieve higher scores than Zero-shot. This indicates that the ground-truth code for questions correctly answered by Zero-shot tends to have fewer lines compared to those answered by other techniques. This finding suggests that selecting PETs based on the code complexity score could be an effective approach that can support our proposal of incorporating code complexity into \tool.

\begin{tcolorbox}[boxrule=0.5pt, colback=gray!10, arc=4pt,left=6pt,right=6pt,top=6pt,bottom=6pt,boxsep=0pt]
\textbf{Finding 1:} In RQ1, we identified two relationships that can inform our design decision of \tool: task types and task complexity. We also include two baseline approaches for selecting appropriate PETs: choosing techniques based on types of tasks or selecting them according to code complexity scores. However, both of these baselines will require additional manual labeling.
\end{tcolorbox}

\begin{table*}[t]
    \centering
    \caption{
    Pass@1 accuracy and token usage were evaluated on benchmark datasets using \tool and twelve baselines, including nine prompting techniques and four selection baselines across different models. Prompt engineering techniques marked with *, indicate that these techniques require iterative rounds to arrive at the answer for each instance. Acc refers to the Pass@1 accuracy, while \#Token represents the average token usage. The highest accuracy scores and the lowest token usage for each dataset and model are highlighted in bold.
    }
    \label{tab:RQ2_ACC}
    \resizebox{0.9\linewidth}{!}{
    \begin{tabular}{l l@{}r l@{}r l@{}r l@{}r}
        \toprule
         LLM &  \multicolumn{4}{c}{GPT-3.5 Turbo} &  \multicolumn{4}{c}{GPT-4o}\\
         
         \cmidrule(lr){1-1} \cmidrule(lr){2-5} \cmidrule(lr){6-9}
         
         Dataset
         &\multicolumn{2}{c}{MBPP}
         &\multicolumn{2}{c}{HumanEval}
         &\multicolumn{2}{c}{MBPP}
         &\multicolumn{2}{c}{HumanEval} \\

         \cmidrule(lr){1-1} \cmidrule(lr){2-3} \cmidrule(lr){4-5} \cmidrule(lr){6-7} \cmidrule(lr){8-9}

         Metrics
         & Acc 
         & \#Token& Acc
         & \#Token& Acc
         & \#Token& Acc
         & \#Token\\
         
         \cmidrule(lr){1-1} \cmidrule(lr){2-5} \cmidrule(lr){6-9}
         
         Zero-shot &  48.2& \textbf{99}& 63.4& 208&53.8& \textbf{114}& 79.9& 263\\
         
         Zero-shot CoT&  39.5&107& 65.9& 230&52.7& 135& 83.0& 308\\
         
         Few-shot &  47.9& 628& 54.3& 795&51.7& 646& 79.9& 835\\
         
         Few-shot CoT&  47.1& 899& \textbf{70.7}& 1142& 49.7& 951& 83.5& 1191\\
         
         Persona& 47.7&127& 68.3& 251&52.4& 143& 83.0& 345\\

         \cmidrule(lr){1-1} \cmidrule(lr){2-5} \cmidrule(lr){6-9}
         
         Self-planning*& 46.7& 849& 62.8& 1365& 49.3& 1686& 73.2&2006\\
         
         Self-refine*& 29.2& 908& 11.6& 1012& 48.3& 1405& 54.9&1731\\
         
         Progressive Hint*& 47.4& 451& 65.2& 882& 52.1& 522& 77.4&1151\\
         
         Self-debug*& 65.3& 3049& 59.1& 3040& 67.6& 4935& 78.7&5518\\

         \cmidrule(lr){1-1} \cmidrule(lr){2-5} \cmidrule(lr){6-9}
         
         Random Selection& 42.7& 642& 57.4& 852& 48.7& 936& 70.7& 1111\\
         
         Category Selection& 50.4& 264& 65.2& \textbf{171}& 55.7& 355& 79.9& \textbf{240}\\
         
         \tool W/o CL& 48.2& 99& 63.4& 208& 53.8& 114& 79.9&263\\
         
         \tool& \textbf{65.6}& 2647& \textbf{70.7}& 409& \textbf{68.2}& 4657& \textbf{85.4}& 300\\
         
        \cmidrule(lr){1-1} \cmidrule(lr){2-5} \cmidrule(lr){6-9}
        
        Average & 48.1& 889& 59.6& 863& 54.2& 1374& 77.5& 1250\\
        \bottomrule
    \end{tabular}
    }
    \vspace{-5pt}
\end{table*}
\subsection{RQ2. How do \tool compare  to single PETs and baselines?} \label{RQ2}
In RQ2, we compare \tool with the individual PETs and our two selected baselines.
Table~\ref{tab:RQ2_ACC} presents the pass@1 accuracy and token usage on the MBPP and HumanEval datasets for nine individual PETs, as well as various PET selection approaches, using GPT-3.5 Turbo and GPT-4o. PETs marked with a star, such as Self-planning, indicate that these techniques require iterative rounds to arrive at the answer for each instance. The `Random Selection' row represents a baseline approach where one of the nine PETs is randomly chosen as the most appropriate for each instance. The overall accuracy and token usage are then calculated based on the selected technique. As we mentioned in RQ1.1,
`Category Selection' is the baseline that randomly selects one of the nine techniques based on the probability of each technique being the most appropriate for a given task, as determined by the ranking score mentioned in Section~\ref{ground-truth}. For example, if the probability of Zero-shot being the most appropriate technique for Algorithm Design is 60\% (i.e., among all the questions correctly answered by the language model, Zero-shot is the most appropriate technique for 60\% of them), then Zero-shot will have a 60\% chance of being selected for questions categorized under Algorithm Design. For the `\tool W/o CL' row, we train the selection model using the original CodeBERT without contrastive learning which does not incorporate the complexity measure. For the `\tool' row, we present the results of selecting PETs based on the output of the selection model.

On the MBPP dataset, \tool achieves 65.6\% accuracy with GPT-3.5 Turbo, which is 0.3\% higher than the best accuracy achieved by Self-debug, a technique that applies the same method across all instances. Furthermore, \tool uses approximately 13\% fewer tokens while achieving higher accuracy compared to Self-debug. This indicates that \tool can effectively identify instances that are simple enough for language models to generate correct code using basic techniques. A similar result is observed when running experiments with GPT-4o, where \tool's accuracy is 0.6\% higher than using only Self-debug, while also utilizing fewer tokens. On the HumanEval dataset, \tool achieves the same accuracy as Few-shot CoT but with 64.2\% fewer tokens when using GPT-3.5 Turbo. With GPT-4o, \tool achieves an accuracy of 85.4\%, which is 1.9\% higher than the best accuracy of the other techniques, while also saving up to 74.8\% fewer tokens. 

Although the Category Selection method does not achieve the highest accuracy, it remains at least the third-best approach among all the baselines, with the lowest token usage when applied to the HumanEval dataset. This indicates that knowing the task category partially helps 
in selecting the optimal PET.

The original CodeBERT 
without contrastive learning incorporating problem complexity does not help the selection model consistently choose the appropriate techniques. Instead, it repeatedly selects Zero-shot, as it often appears to be the best technique among all the options. This result suggests that contrastive learning effectively clusters questions of similar complexity in the embedding space, and is essential in enabling the selection model to accurately choose the optimal PET.

Complex PETs such as Self-debug, which require multiple rounds with language models, may not always be the best choice for all questions. For instance, aside from \tool, while Self-debug performs best on the MBPP dataset, it falls short on the HumanEval dataset, where simpler techniques like Few-shot CoT achieve the highest accuracy. This result provides more examples which support the claim that
applying complex techniques to simpler questions can sometimes result in incorrect answers. 
With \tool, we can identify instances that are simple enough to not require complex techniques, while still generating the correct answers with fewer tokens.

\begin{tcolorbox}[boxrule=0.5pt, colback=gray!10, arc=4pt,left=6pt,right=6pt,top=6pt,bottom=6pt,boxsep=0pt]
\textbf{Finding 2:} Overall, \tool outperforms other baseline approaches across different versions of GPT on both datasets, achieving comparable or up to 1.9\% of accuracy improvement with up to 74.8\% fewer tokens. Compared to other baselines, \tool effectively selects the appropriate techniques based on embeddings adjusted by the contrastive learning CodeBERT model. 
\end{tcolorbox}

\begin{table*}[t]
    \centering
    \caption{
    The ranking effectiveness of selection methods measured with MRR and nDCG metrics
    }
    \label{tab:metrics}
    \resizebox{0.9\linewidth}{!}{
    \begin{tabular}{l l@{ }r l@{ }r l@{ }r l@{ }r}
        \toprule
         LLM &  \multicolumn{4}{c}{GPT-3.5 Turbo} &  \multicolumn{4}{c}{GPT-4o}\\
         
         \cmidrule(lr){1-1} \cmidrule(lr){2-5} \cmidrule(lr){6-9}
         
         Dataset
         &\multicolumn{2}{c}{MBPP}
         &\multicolumn{2}{c}{HumanEval}
         &\multicolumn{2}{c}{MBPP}
         &\multicolumn{2}{c}{HumanEval} \\

         \cmidrule(lr){1-1} \cmidrule(lr){2-3} \cmidrule(lr){4-5} \cmidrule(lr){6-7} \cmidrule(lr){8-9}

         Metrics& MRR& nDCG& MRR& nDCG& MRR& nDCG& MRR& nDCG\\

        \cmidrule(lr){1-1} \cmidrule(lr){2-5} \cmidrule(lr){6-9}
         
         Random Selection& 0.5218& 0.5522& 0.5643& 0.6178& 0.5215& 0.5159& 0.5054& 0.7057\\
         
         Category Selection& 0.5832& 0.6099& 0.6199& 0.6929& 0.6180& 0.5753& 0.6231& 0.7652\\
        
         \tool W/o CL& \textbf{0.7560}& 0.5780& \textbf{0.8638}& 0.6800& \textbf{0.8638}& 0.5588& \textbf{0.8954}&0.7538\\
         
         \tool& 0.5756& \textbf{0.6948}& 0.6186& \textbf{0.7270}& 0.5648& \textbf{0.6840}& 0.6027& \textbf{0.8269}\\
         
        \cmidrule(lr){1-1} \cmidrule(lr){2-5} \cmidrule(lr){6-9}
        
        Average & 0.6092& 0.6087& 0.6667& 0.6794& 0.6420& 0.5835& 0.6567& 0.7629\\
        \bottomrule
    \end{tabular}
    }
    \vspace{-5pt}
\end{table*}
\subsection{RQ3. How is \tool able to select an appropriate technique for each query?} \label{RQ3}
In this section, we perform quantitative and qualitative analyses to assess \tool's ability to select the most appropriate technique for each question.

\subsubsection{Quantitative Analysis} As mentioned in Experimental Setup, we utilize two metrics, Mean Reciprocal Rank (MRR) and Normalized Discounted Cumulative Gain (nDCG), to evaluate \tool's recommendation ability. In Table~\ref{tab:metrics} we present various selection methods' effectiveness measured by MRR and nDCG.
Since we applied 5-fold cross-validation
the MRR and nDCG values are the average results from the test set across five folds. 

Without contrastive learning, \tool W/o CL
achieves a high MRR value across all experiments. This occurs because the selection model consistently chooses Zero-shot as the appropriate technique. As a result, \tool  W/o CL tends to perform well in MRR, since Zero-shot is often the most suitable technique for questions it answers correctly. However, a higher MRR score does not necessarily indicate that the best technique is selected for every instance. It simply means that for the instances where the selected technique provides a correct answer, the chosen method is likely one of the top-performing options. This is further demonstrated in Table~\ref{tab:RQ2_ACC}, where \tool without contrastive learning does not achieve the highest accuracy but often uses fewer tokens than other techniques.


 On the other hand, \tool consistently achieves the highest performance with respect to
 nDCG metric.
 This indicates that it can reliably select techniques that lead to correct answers. Although \tool falls short on the MRR metric, meaning it doesn't always choose the most appropriate technique for every instance, the selected PET still generates the correct code that passes all test cases.
 This is evidenced in Table~\ref{tab:RQ2_ACC}, where \tool outperforms other approaches in terms of accuracy across all experiments. This result indicates that
 \tool is effective in selecting the correct technique that is capable of generating the correct code.

\begin{table}
    \centering
    \caption{Selection result of \tool for example instances. \ding{55} indicates the technique answer correctly on the question, while \ding{51} indicates it answer incorrectly.
    }
    \label{tab:discussion}
    \resizebox{0.9\linewidth}{!}{
    \begin{tabular}{lp{6cm}ccc}
    \toprule
          &Techniques&  Zero-shot& Self-debug &\tool\\
    \midrule
          \multirow{2}{*}{1}&Write a function to find the \textbf{nested list} elements which are present in another list.& \multirow{2}{*}{\ding{55}}& \multirow{2}{*}{\ding{51}}&\multirow{2}{*}{Self-debug}\\ 
    \midrule
          \multirow{2}{*}{2}&Write a function to concatenate the given two tuples to a \textbf{nested tuple}.& \multirow{2}{*}{\ding{55}}& \multirow{2}{*}{\ding{51}}&\multirow{2}{*}{Self-debug}\\
    \midrule
          \multirow{2}{*}{3}&Write a function to check if a \textbf{nested list} is a subset of another \textbf{nested list}.& \multirow{2}{*}{\ding{55}}& \multirow{2}{*}{\ding{51}}&\multirow{2}{*}{Self-debug}\\
    \midrule
          \multirow{2}{*}{4}&Write a function to find the maximum of nth column from \textbf{the given tuple list}.& \multirow{2}{*}{\ding{51}}& \multirow{2}{*}{\ding{55}}&\multirow{2}{*}{Zero-shot}\\
    \midrule
          \multirow{3}{*}{5}&Write a function to find frequency of the elements in \textbf{a given list} of lists using collections module.& \multirow{3}{*}{\ding{51}}& \multirow{3}{*}{\ding{55}}&\multirow{3}{*}{Zero-shot}\\
    \bottomrule
         
    \end{tabular}
    }
    
\end{table}
\subsubsection{Qualitative Analysis} 
This section aims to
provide some additional support for
the experimental results by analyzing the queries that were only answered correctly by Zero-shot (our most simple PET) and successfully selected by \tool. Conversely, we also examined the queries that were only answered correctly by Self-debug (our most complex PET) and were likewise successfully selected by \tool. 
The purpose of these analyses is to provide additional examples that explain the reason why \tool is successful in selecting the correct PET in the previous experiments.

Table~\ref{tab:discussion} lists some example instances in the MBPP dataset. For instance, 
questions containing the term `nested' (numbers 1-3 in table~\ref{tab:discussion}) will likely require complex code as it will likely 
involves iterative loops. 
Complex PETs such as Self-debug are more likely to generate the correct answer.
while basic techniques such as Zero-shot tend to answer incorrectly.
\tool successfully selects the appropriate technique between Zero-shot and Self-debug, indicating that it learns to recognize such keywords in the queries. By placing sentences containing the word `nested' closer together in the embedding space, \tool is able to classify them and select the correct PETs. 

In contrast, sentences that do not contain specific keywords are pushed further away from those that do. As a result, \tool will select relatively basic techniques for those questions. For example, 
the queries 4-5 in table~\ref{tab:discussion}
are also related to the List Processing tasks, they only require a single loop to solve. In this case, Zero-shot is a more appropriate PET
while Self-debug is too complex and sub-optimal. 
Since those questions do not contain the specific keywords that indicate complex problems (e.g., nested), \tool selects Zero-shot instead of Self-debug as the appropriate technique.
The above examples demonstrate that \tool can effectively select the appropriate technique based on code complexity predictions derived from keywords in the queries with the help of contrastive learning. 
By selecting simpler PET when appropriate, \tool not only performs well in all cases but also reduces the overall number of tokens required when compared to complex state-of-the-art PETs such as Self-debug.

\begin{tcolorbox}[boxrule=0.5pt, colback=gray!10, arc=4pt,left=6pt,right=6pt,top=6pt,bottom=6pt,boxsep=0pt]
\textbf{Finding 3:} Through quantitative analysis, we found that while \tool does not always select the most efficient technique in terms of token usage, it still manages to provide correct answers by choosing techniques that are capable of generating the correct code. Additionally, qualitative analysis revealed that \tool's improvement over the best individual PET can be explained by its ability to select
simpler
PET when appropriate which reduces token usage while maintaining a high generated code passing rate.
\end{tcolorbox}

\section{Related work}
\subsection{Code Complexity Prediction}
Code complexity prediction has emerged as a key area of focus in recent research, with various approaches leveraging machine learning and deep learning techniques. A notable advancement is the application of deep learning models, such as hierarchical Transformers, which process method-level code snippets and aggregate them into class-level embeddings~\cite{jeon2023deep}. These models excel in handling longer code sequences, surpassing previous methods through advanced multi-level pre-training objectives that enhance the model's understanding of complexity-related features. Additionally, studies have explored the effectiveness of GPT-3-based models like GitHub Copilot, highlighting both their strengths and limitations in zero-shot complexity prediction~\cite{siddiq2023zero}. While Copilot performs well with linear complexities, specialized deep learning models demonstrate superior overall accuracy.

In contrast, \tool diverges from these methods by not concentrating on code complexity prediction directly from natural language queries. Instead, we employ complexity prediction as an intermediate step to determine the appropriate prompting techniques for answering natural language questions.

\subsection{Automated Prompt Engineering}
Automated prompt engineering is an emerging method to adapt large language models (LLMs) for specific tasks by optimizing prompts without altering the model's core parameters. Techniques like AutoPrompt~\cite{shin2020autoprompt} use gradient-guided search to create prompts for tasks such as sentiment analysis and natural language inference, achieving results comparable to state-of-the-art models without additional fine-tuning. Methods such as prompt tuning~\cite{lester2021power} and prefix-tuning~\cite{li2021prefix} further improve model efficiency by learning task-specific prompts while keeping the language model frozen, significantly reducing the number of tunable parameters. Additionally, approaches like Prompt-OIRL~\cite{sun2023offline} optimize arithmetic reasoning through offline inverse reinforcement learning, offering cost-effective and scalable prompt recommendations. Although these automated prompt engineering approaches optimize the prompt without executing large language models (LLMs), they do not account for the iterative interaction with the models. 

However, these automated prompt engineering methods focus on optimizing a single prompt without accounting for iterative interactions with LLMs throughout the process. In contrast, \tool addresses this limitation by incorporating iterative interaction techniques to select the most suitable prompting strategies for code generation tasks.
\section{Threats to validity}
\subsection{Internal validity}
Our code has been thoroughly reviewed to ensure the implementation is correct, and we have confirmed that the questions in the testing dataset are not present in the question base. We also carefully craft our prompts for each prompting technique, adhering closely to the guidelines outlined in the original paper for each method. However, the way prompts and examples are crafted may influence the performance of each technique, which in turn can affect the results of \tool.
\subsection{External validity}
In our experiment, we use two of the most widely recognized benchmark datasets for code generation, MBPP and HumanEval, to demonstrate the effectiveness of \tool, which is primarily designed for Python programming. The performance of \tool can be different on prompting technique selection for other programming languages. In addition, we incorporate nine fundamental prompting techniques and five representative code complexity metrics across two datasets in our experiments. \tool may perform differently with additional techniques, metrics, and data points. Future work is needed to assess the performance of \tool using a broader range of techniques, metrics, and datasets.
\subsection{Construct validity}
We use MRR, nDCG, pass@k, and token usage calculated by the Tiktoken package to measure the performance of \tool. Our approach may have different performance under other metrics. In this work, we assume that code generation questions with similar code complexity scores are semantically equivalent when contrastively training our CodeBERT-based sentence embeddings. Future research is needed to validate this assumption using different metrics or features.
\section{Conclusion}
In this paper, we introduced \tool, a novel system designed to automatically select appropriate prompt engineering techniques (PETs) for code generation tasks based on code complexity predictions. By leveraging contrastive learning and a CodeBERT-based sentence embedding model, \tool effectively identifies simpler questions and applies suitable techniques, achieving comparable or higher accuracy with fewer tokens. Our evaluation of the MBPP and HumanEval datasets demonstrates that \tool not only enhances performance but also reduces computational costs. Future work will focus on refining the model and exploring its application to other domains.

\section{Data Availability}
We release our code and data through the following link: \\ \url{https://anonymous.4open.science/r/Prompt-Selection-B47F}.

\newpage
\bibliographystyle{ACM-Reference-Format}
\bibliography{refs}


\begin{thebibliography}{58}


\ifx \showCODEN    \undefined \def \showCODEN     #1{\unskip}     \fi
\ifx \showDOI      \undefined \def \showDOI       #1{#1}\fi
\ifx \showISBNx    \undefined \def \showISBNx     #1{\unskip}     \fi
\ifx \showISBNxiii \undefined \def \showISBNxiii  #1{\unskip}     \fi
\ifx \showISSN     \undefined \def \showISSN      #1{\unskip}     \fi
\ifx \showLCCN     \undefined \def \showLCCN      #1{\unskip}     \fi
\ifx \shownote     \undefined \def \shownote      #1{#1}          \fi
\ifx \showarticletitle \undefined \def \showarticletitle #1{#1}   \fi
\ifx \showURL      \undefined \def \showURL       {\relax}        \fi
\providecommand\bibfield[2]{#2}
\providecommand\bibinfo[2]{#2}
\providecommand\natexlab[1]{#1}
\providecommand\showeprint[2][]{arXiv:#2}

\bibitem[Austin et~al\mbox{.}(2021)]%
        {austin2021program}
\bibfield{author}{\bibinfo{person}{Jacob Austin}, \bibinfo{person}{Augustus Odena}, \bibinfo{person}{Maxwell Nye}, \bibinfo{person}{Maarten Bosma}, \bibinfo{person}{Henryk Michalewski}, \bibinfo{person}{David Dohan}, \bibinfo{person}{Ellen Jiang}, \bibinfo{person}{Carrie Cai}, \bibinfo{person}{Michael Terry}, \bibinfo{person}{Quoc Le}, {et~al\mbox{.}}} \bibinfo{year}{2021}\natexlab{}.
\newblock \showarticletitle{Program synthesis with large language models}.
\newblock \bibinfo{journal}{\emph{arXiv preprint arXiv:2108.07732}} (\bibinfo{year}{2021}).
\newblock


\bibitem[Bredin(2017)]%
        {bredin2017tristounet}
\bibfield{author}{\bibinfo{person}{Herv{\'e} Bredin}.} \bibinfo{year}{2017}\natexlab{}.
\newblock \showarticletitle{Tristounet: triplet loss for speaker turn embedding}. In \bibinfo{booktitle}{\emph{2017 IEEE international conference on acoustics, speech and signal processing (ICASSP)}}. IEEE, \bibinfo{pages}{5430--5434}.
\newblock


\bibitem[Brown(2020)]%
        {brown2020language}
\bibfield{author}{\bibinfo{person}{Tom~B Brown}.} \bibinfo{year}{2020}\natexlab{}.
\newblock \showarticletitle{Language models are few-shot learners}.
\newblock \bibinfo{journal}{\emph{arXiv preprint arXiv:2005.14165}} (\bibinfo{year}{2020}).
\newblock


\bibitem[Campbell(2018)]%
        {campbell2018cognitive}
\bibfield{author}{\bibinfo{person}{G~Ann Campbell}.} \bibinfo{year}{2018}\natexlab{}.
\newblock \showarticletitle{Cognitive Complexity-A new way of measuring understandability}.
\newblock \bibinfo{journal}{\emph{SonarSource SA}} (\bibinfo{year}{2018}), \bibinfo{pages}{10}.
\newblock


\bibitem[Chen et~al\mbox{.}(2021)]%
        {chen2021evaluating}
\bibfield{author}{\bibinfo{person}{Mark Chen}, \bibinfo{person}{Jerry Tworek}, \bibinfo{person}{Heewoo Jun}, \bibinfo{person}{Qiming Yuan}, \bibinfo{person}{Henrique Ponde De~Oliveira Pinto}, \bibinfo{person}{Jared Kaplan}, \bibinfo{person}{Harri Edwards}, \bibinfo{person}{Yuri Burda}, \bibinfo{person}{Nicholas Joseph}, \bibinfo{person}{Greg Brockman}, {et~al\mbox{.}}} \bibinfo{year}{2021}\natexlab{}.
\newblock \showarticletitle{Evaluating large language models trained on code}.
\newblock \bibinfo{journal}{\emph{arXiv preprint arXiv:2107.03374}} (\bibinfo{year}{2021}).
\newblock


\bibitem[Chen et~al\mbox{.}(2023)]%
        {chen2023teaching}
\bibfield{author}{\bibinfo{person}{Xinyun Chen}, \bibinfo{person}{Maxwell Lin}, \bibinfo{person}{Nathanael Sch{\"a}rli}, {and} \bibinfo{person}{Denny Zhou}.} \bibinfo{year}{2023}\natexlab{}.
\newblock \showarticletitle{Teaching large language models to self-debug}.
\newblock \bibinfo{journal}{\emph{arXiv preprint arXiv:2304.05128}} (\bibinfo{year}{2023}).
\newblock


\bibitem[Chiang and Lee(2024)]%
        {chiang2024over}
\bibfield{author}{\bibinfo{person}{Cheng-Han Chiang} {and} \bibinfo{person}{Hung-Yi Lee}.} \bibinfo{year}{2024}\natexlab{}.
\newblock \showarticletitle{Over-Reasoning and Redundant Calculation of Large Language Models}. In \bibinfo{booktitle}{\emph{Proceedings of the 18th Conference of the European Chapter of the Association for Computational Linguistics (Volume 2: Short Papers)}}. \bibinfo{pages}{161--169}.
\newblock


\bibitem[Dang et~al\mbox{.}(2022)]%
        {dang2022prompt}
\bibfield{author}{\bibinfo{person}{Hai Dang}, \bibinfo{person}{Lukas Mecke}, \bibinfo{person}{Florian Lehmann}, \bibinfo{person}{Sven Goller}, {and} \bibinfo{person}{Daniel Buschek}.} \bibinfo{year}{2022}\natexlab{}.
\newblock \bibinfo{title}{How to Prompt? Opportunities and Challenges of Zero- and Few-Shot Learning for Human-AI Interaction in Creative Applications of Generative Models}.
\newblock
\newblock
\showeprint[arxiv]{2209.01390}~[cs.HC]


\bibitem[Do et~al\mbox{.}(2024)]%
        {do2024automatic}
\bibfield{author}{\bibinfo{person}{Viet-Tung Do}, \bibinfo{person}{Van-Khanh Hoang}, \bibinfo{person}{Duy-Hung Nguyen}, \bibinfo{person}{Shahab Sabahi}, \bibinfo{person}{Jeff Yang}, \bibinfo{person}{Hajime Hotta}, \bibinfo{person}{Minh-Tien Nguyen}, {and} \bibinfo{person}{Hung Le}.} \bibinfo{year}{2024}\natexlab{}.
\newblock \showarticletitle{Automatic Prompt Selection for Large Language Models}.
\newblock \bibinfo{journal}{\emph{arXiv preprint arXiv:2404.02717}} (\bibinfo{year}{2024}).
\newblock


\bibitem[Ebert et~al\mbox{.}(2016)]%
        {ebert2016cyclomatic}
\bibfield{author}{\bibinfo{person}{Christof Ebert}, \bibinfo{person}{James Cain}, \bibinfo{person}{Giuliano Antoniol}, \bibinfo{person}{Steve Counsell}, {and} \bibinfo{person}{Phillip Laplante}.} \bibinfo{year}{2016}\natexlab{}.
\newblock \showarticletitle{Cyclomatic complexity}.
\newblock \bibinfo{journal}{\emph{IEEE software}} \bibinfo{volume}{33}, \bibinfo{number}{6} (\bibinfo{year}{2016}), \bibinfo{pages}{27--29}.
\newblock


\bibitem[Feng and Chen(2024)]%
        {feng2024prompting}
\bibfield{author}{\bibinfo{person}{Sidong Feng} {and} \bibinfo{person}{Chunyang Chen}.} \bibinfo{year}{2024}\natexlab{}.
\newblock \showarticletitle{Prompting is all you need: Automated android bug replay with large language models}. In \bibinfo{booktitle}{\emph{Proceedings of the 46th IEEE/ACM International Conference on Software Engineering}}. \bibinfo{pages}{1--13}.
\newblock


\bibitem[Feng et~al\mbox{.}(2020)]%
        {feng2020codebert}
\bibfield{author}{\bibinfo{person}{Zhangyin Feng}, \bibinfo{person}{Daya Guo}, \bibinfo{person}{Duyu Tang}, \bibinfo{person}{Nan Duan}, \bibinfo{person}{Xiaocheng Feng}, \bibinfo{person}{Ming Gong}, \bibinfo{person}{Linjun Shou}, \bibinfo{person}{Bing Qin}, \bibinfo{person}{Ting Liu}, \bibinfo{person}{Daxin Jiang}, {et~al\mbox{.}}} \bibinfo{year}{2020}\natexlab{}.
\newblock \showarticletitle{Codebert: A pre-trained model for programming and natural languages}.
\newblock \bibinfo{journal}{\emph{arXiv preprint arXiv:2002.08155}} (\bibinfo{year}{2020}).
\newblock


\bibitem[Hariprasad et~al\mbox{.}(2017)]%
        {hariprasad2017software}
\bibfield{author}{\bibinfo{person}{T Hariprasad}, \bibinfo{person}{G Vidhyagaran}, \bibinfo{person}{K Seenu}, {and} \bibinfo{person}{Chandrasegar Thirumalai}.} \bibinfo{year}{2017}\natexlab{}.
\newblock \showarticletitle{Software complexity analysis using halstead metrics}. In \bibinfo{booktitle}{\emph{2017 International Conference on Trends in Electronics and Informatics (ICEI)}}. IEEE, \bibinfo{pages}{1109--1113}.
\newblock


\bibitem[Hoffer and Ailon(2015)]%
        {hoffer2015deep}
\bibfield{author}{\bibinfo{person}{Elad Hoffer} {and} \bibinfo{person}{Nir Ailon}.} \bibinfo{year}{2015}\natexlab{}.
\newblock \showarticletitle{Deep metric learning using triplet network}. In \bibinfo{booktitle}{\emph{Similarity-based pattern recognition: third international workshop, SIMBAD 2015, Copenhagen, Denmark, October 12-14, 2015. Proceedings 3}}. Springer, \bibinfo{pages}{84--92}.
\newblock


\bibitem[Hossain et~al\mbox{.}(2024)]%
        {hossain2024deep}
\bibfield{author}{\bibinfo{person}{Soneya~Binta Hossain}, \bibinfo{person}{Nan Jiang}, \bibinfo{person}{Qiang Zhou}, \bibinfo{person}{Xiaopeng Li}, \bibinfo{person}{Wen-Hao Chiang}, \bibinfo{person}{Yingjun Lyu}, \bibinfo{person}{Hoan Nguyen}, {and} \bibinfo{person}{Omer Tripp}.} \bibinfo{year}{2024}\natexlab{}.
\newblock \showarticletitle{A deep dive into large language models for automated bug localization and repair}.
\newblock \bibinfo{journal}{\emph{Proceedings of the ACM on Software Engineering}} \bibinfo{volume}{1}, \bibinfo{number}{FSE} (\bibinfo{year}{2024}), \bibinfo{pages}{1471--1493}.
\newblock


\bibitem[Huang et~al\mbox{.}(2023)]%
        {huang2023large}
\bibfield{author}{\bibinfo{person}{Jie Huang}, \bibinfo{person}{Xinyun Chen}, \bibinfo{person}{Swaroop Mishra}, \bibinfo{person}{Huaixiu~Steven Zheng}, \bibinfo{person}{Adams~Wei Yu}, \bibinfo{person}{Xinying Song}, {and} \bibinfo{person}{Denny Zhou}.} \bibinfo{year}{2023}\natexlab{}.
\newblock \showarticletitle{Large language models cannot self-correct reasoning yet}.
\newblock \bibinfo{journal}{\emph{arXiv preprint arXiv:2310.01798}} (\bibinfo{year}{2023}).
\newblock


\bibitem[J{\"a}rvelin and Kek{\"a}l{\"a}inen(2002)]%
        {jarvelin2002cumulated}
\bibfield{author}{\bibinfo{person}{Kalervo J{\"a}rvelin} {and} \bibinfo{person}{Jaana Kek{\"a}l{\"a}inen}.} \bibinfo{year}{2002}\natexlab{}.
\newblock \showarticletitle{Cumulated gain-based evaluation of IR techniques}.
\newblock \bibinfo{journal}{\emph{ACM Transactions on Information Systems (TOIS)}} \bibinfo{volume}{20}, \bibinfo{number}{4} (\bibinfo{year}{2002}), \bibinfo{pages}{422--446}.
\newblock


\bibitem[Jeon et~al\mbox{.}(2023)]%
        {jeon2023deep}
\bibfield{author}{\bibinfo{person}{Mingi Jeon}, \bibinfo{person}{Seung-yeop Baik}, \bibinfo{person}{Joonghyuk Hahn}, \bibinfo{person}{Yo-Sub Han}, {and} \bibinfo{person}{Sang-Ki Ko}.} \bibinfo{year}{2023}\natexlab{}.
\newblock \showarticletitle{Deep learning-based source code complexity prediction}.
\newblock  (\bibinfo{year}{2023}).
\newblock


\bibitem[Jiang et~al\mbox{.}(2024)]%
        {jiang2024survey}
\bibfield{author}{\bibinfo{person}{Juyong Jiang}, \bibinfo{person}{Fan Wang}, \bibinfo{person}{Jiasi Shen}, \bibinfo{person}{Sungju Kim}, {and} \bibinfo{person}{Sunghun Kim}.} \bibinfo{year}{2024}\natexlab{}.
\newblock \showarticletitle{A Survey on Large Language Models for Code Generation}.
\newblock \bibinfo{journal}{\emph{arXiv preprint arXiv:2406.00515}} (\bibinfo{year}{2024}).
\newblock


\bibitem[Jiang et~al\mbox{.}(2023)]%
        {jiang2023self}
\bibfield{author}{\bibinfo{person}{Xue Jiang}, \bibinfo{person}{Yihong Dong}, \bibinfo{person}{Lecheng Wang}, \bibinfo{person}{Fang Zheng}, \bibinfo{person}{Qiwei Shang}, \bibinfo{person}{Ge Li}, \bibinfo{person}{Zhi Jin}, {and} \bibinfo{person}{Wenpin Jiao}.} \bibinfo{year}{2023}\natexlab{}.
\newblock \showarticletitle{Self-planning Code Generation with Large Language Models}.
\newblock \bibinfo{journal}{\emph{ACM Transactions on Software Engineering and Methodology}} (\bibinfo{year}{2023}).
\newblock


\bibitem[Jiang et~al\mbox{.}(2008)]%
        {jiang2008comparing}
\bibfield{author}{\bibinfo{person}{Yue Jiang}, \bibinfo{person}{Bojan Cuki}, \bibinfo{person}{Tim Menzies}, {and} \bibinfo{person}{Nick Bartlow}.} \bibinfo{year}{2008}\natexlab{}.
\newblock \showarticletitle{Comparing design and code metrics for software quality prediction}. In \bibinfo{booktitle}{\emph{Proceedings of the 4th international workshop on Predictor models in software engineering}}. \bibinfo{pages}{11--18}.
\newblock


\bibitem[Khosla et~al\mbox{.}(2020)]%
        {khosla2020supervised}
\bibfield{author}{\bibinfo{person}{Prannay Khosla}, \bibinfo{person}{Piotr Teterwak}, \bibinfo{person}{Chen Wang}, \bibinfo{person}{Aaron Sarna}, \bibinfo{person}{Yonglong Tian}, \bibinfo{person}{Phillip Isola}, \bibinfo{person}{Aaron Maschinot}, \bibinfo{person}{Ce Liu}, {and} \bibinfo{person}{Dilip Krishnan}.} \bibinfo{year}{2020}\natexlab{}.
\newblock \showarticletitle{Supervised contrastive learning}.
\newblock \bibinfo{journal}{\emph{Advances in neural information processing systems}}  \bibinfo{volume}{33} (\bibinfo{year}{2020}), \bibinfo{pages}{18661--18673}.
\newblock


\bibitem[Khot et~al\mbox{.}(2022)]%
        {khot2022decomposed}
\bibfield{author}{\bibinfo{person}{Tushar Khot}, \bibinfo{person}{Harsh Trivedi}, \bibinfo{person}{Matthew Finlayson}, \bibinfo{person}{Yao Fu}, \bibinfo{person}{Kyle Richardson}, \bibinfo{person}{Peter Clark}, {and} \bibinfo{person}{Ashish Sabharwal}.} \bibinfo{year}{2022}\natexlab{}.
\newblock \showarticletitle{Decomposed prompting: A modular approach for solving complex tasks}.
\newblock \bibinfo{journal}{\emph{arXiv preprint arXiv:2210.02406}} (\bibinfo{year}{2022}).
\newblock


\bibitem[Kim et~al\mbox{.}(2024)]%
        {kim2024leveraging}
\bibfield{author}{\bibinfo{person}{Myeongsoo Kim}, \bibinfo{person}{Tyler Stennett}, \bibinfo{person}{Dhruv Shah}, \bibinfo{person}{Saurabh Sinha}, {and} \bibinfo{person}{Alessandro Orso}.} \bibinfo{year}{2024}\natexlab{}.
\newblock \showarticletitle{Leveraging large language models to improve REST API testing}. In \bibinfo{booktitle}{\emph{Proceedings of the 2024 ACM/IEEE 44th International Conference on Software Engineering: New Ideas and Emerging Results}}. \bibinfo{pages}{37--41}.
\newblock


\bibitem[Kojima et~al\mbox{.}(2022)]%
        {kojima2022large}
\bibfield{author}{\bibinfo{person}{Takeshi Kojima}, \bibinfo{person}{Shixiang~Shane Gu}, \bibinfo{person}{Machel Reid}, \bibinfo{person}{Yutaka Matsuo}, {and} \bibinfo{person}{Yusuke Iwasawa}.} \bibinfo{year}{2022}\natexlab{}.
\newblock \showarticletitle{Large language models are zero-shot reasoners}.
\newblock \bibinfo{journal}{\emph{Advances in neural information processing systems}}  \bibinfo{volume}{35} (\bibinfo{year}{2022}), \bibinfo{pages}{22199--22213}.
\newblock


\bibitem[Lester et~al\mbox{.}(2021)]%
        {lester2021power}
\bibfield{author}{\bibinfo{person}{Brian Lester}, \bibinfo{person}{Rami Al-Rfou}, {and} \bibinfo{person}{Noah Constant}.} \bibinfo{year}{2021}\natexlab{}.
\newblock \showarticletitle{The power of scale for parameter-efficient prompt tuning}.
\newblock \bibinfo{journal}{\emph{arXiv preprint arXiv:2104.08691}} (\bibinfo{year}{2021}).
\newblock


\bibitem[Li and Liang(2021)]%
        {li2021prefix}
\bibfield{author}{\bibinfo{person}{Xiang~Lisa Li} {and} \bibinfo{person}{Percy Liang}.} \bibinfo{year}{2021}\natexlab{}.
\newblock \showarticletitle{Prefix-tuning: Optimizing continuous prompts for generation}.
\newblock \bibinfo{journal}{\emph{arXiv preprint arXiv:2101.00190}} (\bibinfo{year}{2021}).
\newblock


\bibitem[Liu et~al\mbox{.}(2021)]%
        {liu2021makes}
\bibfield{author}{\bibinfo{person}{Jiachang Liu}, \bibinfo{person}{Dinghan Shen}, \bibinfo{person}{Yizhe Zhang}, \bibinfo{person}{Bill Dolan}, \bibinfo{person}{Lawrence Carin}, {and} \bibinfo{person}{Weizhu Chen}.} \bibinfo{year}{2021}\natexlab{}.
\newblock \showarticletitle{What Makes Good In-Context Examples for GPT-$3 $?}
\newblock \bibinfo{journal}{\emph{arXiv preprint arXiv:2101.06804}} (\bibinfo{year}{2021}).
\newblock


\bibitem[Lu et~al\mbox{.}(2021)]%
        {lu2021fantastically}
\bibfield{author}{\bibinfo{person}{Yao Lu}, \bibinfo{person}{Max Bartolo}, \bibinfo{person}{Alastair Moore}, \bibinfo{person}{Sebastian Riedel}, {and} \bibinfo{person}{Pontus Stenetorp}.} \bibinfo{year}{2021}\natexlab{}.
\newblock \showarticletitle{Fantastically ordered prompts and where to find them: Overcoming few-shot prompt order sensitivity}.
\newblock \bibinfo{journal}{\emph{arXiv preprint arXiv:2104.08786}} (\bibinfo{year}{2021}).
\newblock


\bibitem[Madaan et~al\mbox{.}(2024)]%
        {madaan2024self}
\bibfield{author}{\bibinfo{person}{Aman Madaan}, \bibinfo{person}{Niket Tandon}, \bibinfo{person}{Prakhar Gupta}, \bibinfo{person}{Skyler Hallinan}, \bibinfo{person}{Luyu Gao}, \bibinfo{person}{Sarah Wiegreffe}, \bibinfo{person}{Uri Alon}, \bibinfo{person}{Nouha Dziri}, \bibinfo{person}{Shrimai Prabhumoye}, \bibinfo{person}{Yiming Yang}, {et~al\mbox{.}}} \bibinfo{year}{2024}\natexlab{}.
\newblock \showarticletitle{Self-refine: Iterative refinement with self-feedback}.
\newblock \bibinfo{journal}{\emph{Advances in Neural Information Processing Systems}}  \bibinfo{volume}{36} (\bibinfo{year}{2024}).
\newblock


\bibitem[Nashid et~al\mbox{.}(2023)]%
        {nashid2023retrieval}
\bibfield{author}{\bibinfo{person}{Noor Nashid}, \bibinfo{person}{Mifta Sintaha}, {and} \bibinfo{person}{Ali Mesbah}.} \bibinfo{year}{2023}\natexlab{}.
\newblock \showarticletitle{Retrieval-based prompt selection for code-related few-shot learning}. In \bibinfo{booktitle}{\emph{2023 IEEE/ACM 45th International Conference on Software Engineering (ICSE)}}. IEEE, \bibinfo{pages}{2450--2462}.
\newblock


\bibitem[Oord et~al\mbox{.}(2018)]%
        {oord2018representation}
\bibfield{author}{\bibinfo{person}{Aaron van~den Oord}, \bibinfo{person}{Yazhe Li}, {and} \bibinfo{person}{Oriol Vinyals}.} \bibinfo{year}{2018}\natexlab{}.
\newblock \showarticletitle{Representation learning with contrastive predictive coding}.
\newblock \bibinfo{journal}{\emph{arXiv preprint arXiv:1807.03748}} (\bibinfo{year}{2018}).
\newblock


\bibitem[Perez et~al\mbox{.}(2021)]%
        {perez2021true}
\bibfield{author}{\bibinfo{person}{Ethan Perez}, \bibinfo{person}{Douwe Kiela}, {and} \bibinfo{person}{Kyunghyun Cho}.} \bibinfo{year}{2021}\natexlab{}.
\newblock \showarticletitle{True few-shot learning with language models}.
\newblock \bibinfo{journal}{\emph{Advances in neural information processing systems}}  \bibinfo{volume}{34} (\bibinfo{year}{2021}), \bibinfo{pages}{11054--11070}.
\newblock


\bibitem[Radev et~al\mbox{.}(2002)]%
        {radev2002evaluating}
\bibfield{author}{\bibinfo{person}{Dragomir~R Radev}, \bibinfo{person}{Hong Qi}, \bibinfo{person}{Harris Wu}, {and} \bibinfo{person}{Weiguo Fan}.} \bibinfo{year}{2002}\natexlab{}.
\newblock \showarticletitle{Evaluating web-based question answering systems.}. In \bibinfo{booktitle}{\emph{LREC}}. Citeseer.
\newblock


\bibitem[Rubin et~al\mbox{.}(2021)]%
        {rubin2021learning}
\bibfield{author}{\bibinfo{person}{Ohad Rubin}, \bibinfo{person}{Jonathan Herzig}, {and} \bibinfo{person}{Jonathan Berant}.} \bibinfo{year}{2021}\natexlab{}.
\newblock \showarticletitle{Learning to retrieve prompts for in-context learning}.
\newblock \bibinfo{journal}{\emph{arXiv preprint arXiv:2112.08633}} (\bibinfo{year}{2021}).
\newblock


\bibitem[Sch{\"a}fer et~al\mbox{.}(2023)]%
        {schafer2023empirical}
\bibfield{author}{\bibinfo{person}{Max Sch{\"a}fer}, \bibinfo{person}{Sarah Nadi}, \bibinfo{person}{Aryaz Eghbali}, {and} \bibinfo{person}{Frank Tip}.} \bibinfo{year}{2023}\natexlab{}.
\newblock \showarticletitle{An empirical evaluation of using large language models for automated unit test generation}.
\newblock \bibinfo{journal}{\emph{IEEE Transactions on Software Engineering}} (\bibinfo{year}{2023}).
\newblock


\bibitem[Shin et~al\mbox{.}(2024a)]%
        {shin2023domain}
\bibfield{author}{\bibinfo{person}{Jiho Shin}, \bibinfo{person}{Sepehr Hashtroudi}, \bibinfo{person}{Hadi Hemmati}, {and} \bibinfo{person}{Song Wang}.} \bibinfo{year}{2024}\natexlab{a}.
\newblock \showarticletitle{Domain Adaptation for Code Model-Based Unit Test Case Generation}. In \bibinfo{booktitle}{\emph{Proceedings of the 33rd ACM SIGSOFT International Symposium on Software Testing and Analysis}} (Vienna, Austria) \emph{(\bibinfo{series}{ISSTA 2024})}. \bibinfo{publisher}{Association for Computing Machinery}, \bibinfo{address}{New York, NY, USA}, \bibinfo{pages}{1211–1222}.
\newblock
\showISBNx{9798400706127}
\urldef\tempurl%
\url{https://doi.org/10.1145/3650212.3680354}
\showDOI{\tempurl}


\bibitem[Shin et~al\mbox{.}(2024b)]%
        {shin2024assessing}
\bibfield{author}{\bibinfo{person}{Jiho Shin}, \bibinfo{person}{Hadi Hemmati}, \bibinfo{person}{Moshi Wei}, {and} \bibinfo{person}{Song Wang}.} \bibinfo{year}{2024}\natexlab{b}.
\newblock \showarticletitle{Assessing evaluation metrics for neural test oracle generation}.
\newblock \bibinfo{journal}{\emph{IEEE Transactions on Software Engineering}} (\bibinfo{year}{2024}).
\newblock


\bibitem[Shin and Nam(2021)]%
        {shin2021survey}
\bibfield{author}{\bibinfo{person}{Jiho Shin} {and} \bibinfo{person}{Jaechang Nam}.} \bibinfo{year}{2021}\natexlab{}.
\newblock \showarticletitle{A survey of automatic code generation from natural language}.
\newblock \bibinfo{journal}{\emph{Journal of Information Processing Systems}} \bibinfo{volume}{17}, \bibinfo{number}{3} (\bibinfo{year}{2021}), \bibinfo{pages}{537--555}.
\newblock


\bibitem[Shin et~al\mbox{.}(2023)]%
        {shin2023prompt}
\bibfield{author}{\bibinfo{person}{Jiho Shin}, \bibinfo{person}{Clark Tang}, \bibinfo{person}{Tahmineh Mohati}, \bibinfo{person}{Maleknaz Nayebi}, \bibinfo{person}{Song Wang}, {and} \bibinfo{person}{Hadi Hemmati}.} \bibinfo{year}{2023}\natexlab{}.
\newblock \showarticletitle{Prompt engineering or fine tuning: An empirical assessment of large language models in automated software engineering tasks}.
\newblock \bibinfo{journal}{\emph{arXiv preprint arXiv:2310.10508}} (\bibinfo{year}{2023}).
\newblock


\bibitem[Shin et~al\mbox{.}(2020)]%
        {shin2020autoprompt}
\bibfield{author}{\bibinfo{person}{Taylor Shin}, \bibinfo{person}{Yasaman Razeghi}, \bibinfo{person}{Robert~L Logan~IV}, \bibinfo{person}{Eric Wallace}, {and} \bibinfo{person}{Sameer Singh}.} \bibinfo{year}{2020}\natexlab{}.
\newblock \showarticletitle{Autoprompt: Eliciting knowledge from language models with automatically generated prompts}.
\newblock \bibinfo{journal}{\emph{arXiv preprint arXiv:2010.15980}} (\bibinfo{year}{2020}).
\newblock


\bibitem[Shin and Williams(2008)]%
        {shin2008empirical}
\bibfield{author}{\bibinfo{person}{Yonghee Shin} {and} \bibinfo{person}{Laurie Williams}.} \bibinfo{year}{2008}\natexlab{}.
\newblock \showarticletitle{An empirical model to predict security vulnerabilities using code complexity metrics}. In \bibinfo{booktitle}{\emph{Proceedings of the Second ACM-IEEE international symposium on Empirical software engineering and measurement}}. \bibinfo{pages}{315--317}.
\newblock


\bibitem[Siddiq et~al\mbox{.}(2023)]%
        {siddiq2023zero}
\bibfield{author}{\bibinfo{person}{Mohammed~Latif Siddiq}, \bibinfo{person}{Abdus Samee}, \bibinfo{person}{Sk~Ruhul Azgor}, \bibinfo{person}{Md~Asif Haider}, \bibinfo{person}{Shehabul~Islam Sawraz}, {and} \bibinfo{person}{Joanna~CS Santos}.} \bibinfo{year}{2023}\natexlab{}.
\newblock \showarticletitle{Zero-shot prompting for code complexity prediction using github copilot}. In \bibinfo{booktitle}{\emph{2023 IEEE/ACM 2nd International Workshop on Natural Language-Based Software Engineering (NLBSE)}}. IEEE, \bibinfo{pages}{56--59}.
\newblock


\bibitem[Sun(2023)]%
        {sun2023offline}
\bibfield{author}{\bibinfo{person}{Hao Sun}.} \bibinfo{year}{2023}\natexlab{}.
\newblock \showarticletitle{Offline prompt evaluation and optimization with inverse reinforcement learning}.
\newblock \bibinfo{journal}{\emph{arXiv preprint arXiv:2309.06553}} (\bibinfo{year}{2023}).
\newblock


\bibitem[Tony et~al\mbox{.}(2024)]%
        {tony2024prompting}
\bibfield{author}{\bibinfo{person}{Catherine Tony}, \bibinfo{person}{Nicol{\'a}s E~D{\'\i}az Ferreyra}, \bibinfo{person}{Markus Mutas}, \bibinfo{person}{Salem Dhiff}, {and} \bibinfo{person}{Riccardo Scandariato}.} \bibinfo{year}{2024}\natexlab{}.
\newblock \showarticletitle{Prompting Techniques for Secure Code Generation: A Systematic Investigation}.
\newblock \bibinfo{journal}{\emph{arXiv preprint arXiv:2407.07064}} (\bibinfo{year}{2024}).
\newblock


\bibitem[Voorhees et~al\mbox{.}(1999)]%
        {voorhees1999trec}
\bibfield{author}{\bibinfo{person}{Ellen~M Voorhees} {et~al\mbox{.}}} \bibinfo{year}{1999}\natexlab{}.
\newblock \showarticletitle{The trec-8 question answering track report.}. In \bibinfo{booktitle}{\emph{Trec}}, Vol.~\bibinfo{volume}{99}. \bibinfo{pages}{77--82}.
\newblock


\bibitem[Wei et~al\mbox{.}(2022b)]%
        {wei2022chain}
\bibfield{author}{\bibinfo{person}{Jason Wei}, \bibinfo{person}{Xuezhi Wang}, \bibinfo{person}{Dale Schuurmans}, \bibinfo{person}{Maarten Bosma}, \bibinfo{person}{Fei Xia}, \bibinfo{person}{Ed Chi}, \bibinfo{person}{Quoc~V Le}, \bibinfo{person}{Denny Zhou}, {et~al\mbox{.}}} \bibinfo{year}{2022}\natexlab{b}.
\newblock \showarticletitle{Chain-of-thought prompting elicits reasoning in large language models}.
\newblock \bibinfo{journal}{\emph{Advances in neural information processing systems}}  \bibinfo{volume}{35} (\bibinfo{year}{2022}), \bibinfo{pages}{24824--24837}.
\newblock


\bibitem[Wei et~al\mbox{.}(2022a)]%
        {wei2022clear}
\bibfield{author}{\bibinfo{person}{Moshi Wei}, \bibinfo{person}{Nima~Shiri Harzevili}, \bibinfo{person}{Yuchao Huang}, \bibinfo{person}{Junjie Wang}, {and} \bibinfo{person}{Song Wang}.} \bibinfo{year}{2022}\natexlab{a}.
\newblock \showarticletitle{Clear: contrastive learning for api recommendation}. In \bibinfo{booktitle}{\emph{Proceedings of the 44th International Conference on Software Engineering}}. \bibinfo{pages}{376--387}.
\newblock


\bibitem[Wei et~al\mbox{.}(2023)]%
        {wei2023copiloting}
\bibfield{author}{\bibinfo{person}{Yuxiang Wei}, \bibinfo{person}{Chunqiu~Steven Xia}, {and} \bibinfo{person}{Lingming Zhang}.} \bibinfo{year}{2023}\natexlab{}.
\newblock \showarticletitle{Copiloting the copilots: Fusing large language models with completion engines for automated program repair}. In \bibinfo{booktitle}{\emph{Proceedings of the 31st ACM Joint European Software Engineering Conference and Symposium on the Foundations of Software Engineering}}. \bibinfo{pages}{172--184}.
\newblock


\bibitem[Welker(2001)]%
        {welker2001software}
\bibfield{author}{\bibinfo{person}{Kurt~D Welker}.} \bibinfo{year}{2001}\natexlab{}.
\newblock \showarticletitle{The software maintainability index revisited}.
\newblock \bibinfo{journal}{\emph{CrossTalk}}  \bibinfo{volume}{14} (\bibinfo{year}{2001}), \bibinfo{pages}{18--21}.
\newblock


\bibitem[White et~al\mbox{.}(2023)]%
        {white2023prompt}
\bibfield{author}{\bibinfo{person}{Jules White}, \bibinfo{person}{Quchen Fu}, \bibinfo{person}{Sam Hays}, \bibinfo{person}{Michael Sandborn}, \bibinfo{person}{Carlos Olea}, \bibinfo{person}{Henry Gilbert}, \bibinfo{person}{Ashraf Elnashar}, \bibinfo{person}{Jesse Spencer-Smith}, {and} \bibinfo{person}{Douglas~C Schmidt}.} \bibinfo{year}{2023}\natexlab{}.
\newblock \showarticletitle{A prompt pattern catalog to enhance prompt engineering with chatgpt}.
\newblock \bibinfo{journal}{\emph{arXiv preprint arXiv:2302.11382}} (\bibinfo{year}{2023}).
\newblock


\bibitem[Yang et~al\mbox{.}(2024)]%
        {yang2024buffer}
\bibfield{author}{\bibinfo{person}{Ling Yang}, \bibinfo{person}{Zhaochen Yu}, \bibinfo{person}{Tianjun Zhang}, \bibinfo{person}{Shiyi Cao}, \bibinfo{person}{Minkai Xu}, \bibinfo{person}{Wentao Zhang}, \bibinfo{person}{Joseph~E Gonzalez}, {and} \bibinfo{person}{Bin Cui}.} \bibinfo{year}{2024}\natexlab{}.
\newblock \showarticletitle{Buffer of Thoughts: Thought-Augmented Reasoning with Large Language Models}.
\newblock \bibinfo{journal}{\emph{arXiv preprint arXiv:2406.04271}} (\bibinfo{year}{2024}).
\newblock


\bibitem[Yao et~al\mbox{.}(2024)]%
        {yao2024tree}
\bibfield{author}{\bibinfo{person}{Shunyu Yao}, \bibinfo{person}{Dian Yu}, \bibinfo{person}{Jeffrey Zhao}, \bibinfo{person}{Izhak Shafran}, \bibinfo{person}{Tom Griffiths}, \bibinfo{person}{Yuan Cao}, {and} \bibinfo{person}{Karthik Narasimhan}.} \bibinfo{year}{2024}\natexlab{}.
\newblock \showarticletitle{Tree of thoughts: Deliberate problem solving with large language models}.
\newblock \bibinfo{journal}{\emph{Advances in Neural Information Processing Systems}}  \bibinfo{volume}{36} (\bibinfo{year}{2024}).
\newblock


\bibitem[Zhang et~al\mbox{.}(2007)]%
        {zhang2007predicting}
\bibfield{author}{\bibinfo{person}{Hongyu Zhang}, \bibinfo{person}{Xiuzhen Zhang}, {and} \bibinfo{person}{Ming Gu}.} \bibinfo{year}{2007}\natexlab{}.
\newblock \showarticletitle{Predicting defective software components from code complexity measures}. In \bibinfo{booktitle}{\emph{13th Pacific Rim International Symposium on Dependable Computing (PRDC 2007)}}. IEEE, \bibinfo{pages}{93--96}.
\newblock


\bibitem[Zhao et~al\mbox{.}(2023)]%
        {zhao2023automatic}
\bibfield{author}{\bibinfo{person}{James~Xu Zhao}, \bibinfo{person}{Yuxi Xie}, \bibinfo{person}{Kenji Kawaguchi}, \bibinfo{person}{Junxian He}, {and} \bibinfo{person}{Michael~Qizhe Xie}.} \bibinfo{year}{2023}\natexlab{}.
\newblock \showarticletitle{Automatic model selection with large language models for reasoning}.
\newblock \bibinfo{journal}{\emph{arXiv preprint arXiv:2305.14333}} (\bibinfo{year}{2023}).
\newblock


\bibitem[Zheng et~al\mbox{.}(2023)]%
        {zheng2023progressive}
\bibfield{author}{\bibinfo{person}{Chuanyang Zheng}, \bibinfo{person}{Zhengying Liu}, \bibinfo{person}{Enze Xie}, \bibinfo{person}{Zhenguo Li}, {and} \bibinfo{person}{Yu Li}.} \bibinfo{year}{2023}\natexlab{}.
\newblock \showarticletitle{Progressive-hint prompting improves reasoning in large language models}.
\newblock \bibinfo{journal}{\emph{arXiv preprint arXiv:2304.09797}} (\bibinfo{year}{2023}).
\newblock


\bibitem[Zhou et~al\mbox{.}(2022b)]%
        {zhou2022least}
\bibfield{author}{\bibinfo{person}{Denny Zhou}, \bibinfo{person}{Nathanael Sch{\"a}rli}, \bibinfo{person}{Le Hou}, \bibinfo{person}{Jason Wei}, \bibinfo{person}{Nathan Scales}, \bibinfo{person}{Xuezhi Wang}, \bibinfo{person}{Dale Schuurmans}, \bibinfo{person}{Claire Cui}, \bibinfo{person}{Olivier Bousquet}, \bibinfo{person}{Quoc Le}, {et~al\mbox{.}}} \bibinfo{year}{2022}\natexlab{b}.
\newblock \showarticletitle{Least-to-most prompting enables complex reasoning in large language models}.
\newblock \bibinfo{journal}{\emph{arXiv preprint arXiv:2205.10625}} (\bibinfo{year}{2022}).
\newblock


\bibitem[Zhou et~al\mbox{.}(2022a)]%
        {zhou2022large}
\bibfield{author}{\bibinfo{person}{Yongchao Zhou}, \bibinfo{person}{Andrei~Ioan Muresanu}, \bibinfo{person}{Ziwen Han}, \bibinfo{person}{Keiran Paster}, \bibinfo{person}{Silviu Pitis}, \bibinfo{person}{Harris Chan}, {and} \bibinfo{person}{Jimmy Ba}.} \bibinfo{year}{2022}\natexlab{a}.
\newblock \showarticletitle{Large language models are human-level prompt engineers}.
\newblock \bibinfo{journal}{\emph{arXiv preprint arXiv:2211.01910}} (\bibinfo{year}{2022}).
\newblock


\end{thebibliography}
\end{document}